\documentclass[12pt]{article}
\pdfoutput=1
\def\parn              {  \par\noindent }
\def\parbigskip        {  \par\bigskip  }

\def\ep{\epsilon}

\def\sig{\sigma}

  \def\calO{{\cal O}}

\def\atil{{\tilde{a}}}
\def\btil{{\tilde{b}}}

\def\zbar{{\bar{z}}}
\def\Zbar{\bar{Z}}

\def\Xbar{\bar{X}}

\def\rootof#1   {  \left( #1 \right)^{1/2}  } 
\def\trace      {  \mbox{Tr}\,  }

\def\abs#1      {  \vert #1 \vert  }
\def\ie         {{\it i.e.}\,\,}
\def\evalat#1   {  \left\vert_{#1} \right. }

\def\comma          {\, ,}
\def\period         {\, .}
\def\lsim      {\lower .65ex \hbox{\ $\stackrel{<}{\sim}$\ } }
\def\gsim      {\lower .65ex \hbox{\ $\stackrel{>}{\sim}$\ } }
\def\det       {{\rm det}\, }

\def\bra#1{{\langle #1 | } }
\def\ket#1{{| #1 \rangle } }

\def\matel#1#2#3  {{\langle #1 | #2 | #3 \rangle } }

\def\lrvec#1    {\hbox{$\stackrel{\leftrightarrow}{#1}$}}
\def\lvec#1     {\hbox{$\stackrel{\leftarrow}{#1}$}}

\def\vecii#1#2      {  \left(\begin{array}{c}#1\\#2\end{array}\right)  }
\def\veciii#1#2#3   {  \left(\begin{array}{c}#1\\#2\\#3\end{array}
                     \right)  }
\def\veciv#1#2#3#4  {  \left(\begin{array}{c}#1\\#2\\#3\\#4
                                 \end{array}\right)  }
\def\vecfv#1#2#3#4#5 {  \left(\begin{array}{c}#1\\#2\\#3\\#4\\#5
                                 \end{array}\right)  }

\def\matrixii#1#2#3#4            {  \left(\begin{array}{cc}#1&#2\\#3&#4
                                       \end{array}\right) }
\def\matrixiii#1#2#3#4#5#6#7#8#9 {  \left(\begin{array}{ccc}#1&#2&#3\\
                                     #4&#5&#6\\#7&#8&#9\end{array}
                               \right)  }
\def\mativ#1#2#3#4               {  \left(\begin{array}{cccc}
                                       #1\\#2\\#3\\#4\end{array}\right) }
\def\matv#1#2#3#4#5              {  \left(\begin{array}{ccccc}
                                     #1\\#2\\#3\\#4\\#5\end{array}
                              \right)  }
\def\eqabegin         {  \begin{eqnarray}  }
\def\eqaend           {  \end{eqnarray}  }
\def\nn               {  \nonumber  }
\def\bracetwo#1#2     {  \left\{ \begin{array}{l} #1 \\ #2 \end{array}
                         \right.  }
\def\bracetwocases#1#2#3#4  {   \left\{ \begin{array}{ll} #1 &
                                 \qquad #2 \\
                                 #3 & \qquad #4 \end{array} \right.  }
\def\bracebegin#1     {  \left\{ \begin{array}{#1}   }
\def\braceend         {  \end{array}\right.   }
\def\boxit#1#2      {  \vbox{\hrule\hbox{ \hskip -4.1pt \vrule\kern3pt 
                     \vbox
                    {  \hsize #1 \strut\kern3pt #2 \kern3pt\strut  }
                       \kern3pt  \vrule} \hrule  } }
\def\centerbox#1#2  {  \mbox{  }\par\bigskip  \hfil \boxit{#1}{#2} \hfil
                       \par\bigskip\noindent }
\def\rightbox#1#2   {  \hfill\boxit{#1}{#2}  }
\def\leftbox#1#2    {  \boxit{#1}{#2}  }
\def\fullbox#1      {  \boxit{\textwidth}{#1}  }

\newcommand{\nullify}[1]{}

\def\SUR{SU(2)$_R$ } 
\def\SUL{SU(2)$_L$ } 
\def\SULR{{\rm SU(2)}_L \times {\rm SU(2)}_R }
\def\papertitlepage{\baselineskip 3.5ex \thispagestyle{empty}}
\def\Title#1{\baselineskip 1cm \vspace{1.5cm}\begin{center}
 {\Large\bf #1} \end{center} 
\vspace{0.5cm}}
\def\Authors#1{\begin{center} {\it #1} \end{center}}
\def\Abstract{\vspace{1.0cm}\begin{center} {\large\bf Abstract} 
           \end{center} \par\bigskip}
\def\Komabanumber#1#2#3{\hfill \begin{minipage}{4.2cm} UT-Komaba #1
              \parn #2 
              \parn #3 \end{minipage}}
\renewcommand{\thefootnote}{\fnsymbol{footnote}}

\newcommand{\beqa}{\begin{eqnarray}}
\newcommand{\eeqa}{\end{eqnarray}}
\newcommand{\bea}{\begin{array}}
\newcommand{\eea}{\end{array}}
\newcommand{\beqn}{\begin{equation}}
\newcommand{\eeqn}{\end{equation}}
\usepackage{amssymb}
\usepackage{amsmath}
\usepackage{latexsym}
\usepackage[usenames]{color}
\usepackage{fancybox}
\usepackage{underbracket}
\usepackage{cite}
\definecolor{refkey}{rgb}{0.5,0.5,0}
\definecolor{labelkey}{rgb}{0.5,0.5,0}
\definecolor{citekey}{rgb}{0.5,0.5,0}
\definecolor{darkgreen}{rgb}{0,0.5,0}
\definecolor{darkblue}{cmyk}{0.9,0.9,0,0}
\definecolor{darkred}{rgb}{0.6,0,0.3}
\definecolor{MyRed}{cmyk}{0,1,1,0.15}

\usepackage{ifpdf}
\ifpdf
\usepackage{graphicx} 
\usepackage{epsfig}
\usepackage[setpagesize=false,pagebackref=false, linktocpage, bookmarksopen=true, colorlinks=true, linkcolor=darkblue,citecolor=darkblue,urlcolor=darkblue]{hyperref}
\usepackage{hyperref}
\newcommand{\arXiv}[2]{[\href{http://arxiv.org/abs/#1}{{\tt arXiv:#2}}]}
\newcommand{\hep}[2]{[\href{http://arxiv.org/abs/#1}{{\tt #2}}]}
\def\picture#1#2{\includegraphics[#1]{#2.pdf}}
\else
\usepackage{graphicx}
\newcommand{\arXiv}[2]{[{\tt arXiv:#2}]}
\newcommand{\hep}[2]{[{\tt #2}]}
\def\picture#1#2{\includegraphics[#1]{#2.eps}}
\fi

\newcommand{\tr}{{\rm tr}}

\def\fn#1{\footnote{#1}}
\def\nn{\nonumber}
\def\eqref#1{(\ref{#1})}
\def\comma{\,,}
\def\period{\,.}
\def\figref#1{Figure \ref{#1}}
\def\beq#1{\begin{align}#1\end{align}}
\def\pmatrix#1#2{\left( 
\begin{array}{#1}
#2\end{array} 
\right)}
\def\upspin{\!\uparrow}
\def\downspin{\!\downarrow}
\def\upket{\ket{\!\uparrow}}
\def\downket{\ket{\!\downarrow}}

\def\ewick#1#2{\cunderbracket{$#1$}{ }{$#2$}}

\def\P{{\rm P}}
\def\n{\mathfrak{n}}
\def\bm#1{\text{\boldmath $#1$}}
\def\g{\mathfrak{g}}
\def\s{\bm{1}}
 
\begin{document}
\papertitlepage
\vspace*{-1cm}
\Komabanumber{14-4}{October, 2014}{}
\vspace{-0.8cm}
\Title{Novel construction and the monodromy relation \\
 for  three-point functions  at weak coupling  } 
\Authors{\baselineskip 3ex
{\sc Yoichi Kazama\footnote[2]{\textcolor{black}{\tt kazama@hep1.c.u-tokyo.ac.jp}}, 
 Shota Komatsu\footnote[3]{\textcolor{black}{\tt skomatsu@perimeterinstitute.ca}} and Takuya Nishimura\footnote[4]{\textcolor{black}{\tt tnishimura@hep1.c.u-tokyo.ac.jp}} 
\\ }
\vskip 2ex
   ${}^\dagger\,  {}^\S$ Institute of Physics, University of Tokyo, \\
 Komaba, Meguro-ku, Tokyo 153-8902 Japan \parbigskip\noindent
${}^\ddagger$ Perimeter Institute for Theoretical Physics, \\
\vspace{-0.4cm}
Waterloo, Ontario N2L 2Y5, Canada 
  }
\vspace{-1cm}
\renewcommand{\thefootnote}{\arabic{footnote}}
\numberwithin{equation}{section}
\numberwithin{figure}{section}
\numberwithin{table}{section}
\parskip=0.9ex
\baselineskip 3ex
\Abstract
In this article, we shall develop and formulate two novel viewpoints and properties concerning the three-point functions at weak coupling in the SU(2) sector of the $\mathcal{N}=4$ super Yang-Mills theory. One is a double spin-chain formulation of the spin-chain and the associated new interpretation of the operation of  Wick contraction.  It will be regarded as a skew symmetric pairing which acts as a projection onto a singlet  in the entire SO(4) sector, instead of an inner product in the spin-chain Hilbert space. This formalism allows us to study a class of three-point functions of operators built upon more general spin-chain vacua than the special configuration discussed so far in the literature.  
Furthermore, this new viewpoint has the significant advantage over the conventional method:  In the usual ``tailoring" operation,  the Wick contraction produces inner products between off-shell Bethe states, which cannot be in general converted into simple expressions. In contrast, our procedure directly produces the so-called partial domain wall partition functions, which can 
be expressed as determinants. Using this property, we derive simple determinantal representation for a broader class of three-point functions. The second new property uncovered in this work is the 
non-trivial identity satisfied by the  three-point functions with monodromy operators  inserted. Generically this relation connects three-point functions of different operators and can be regarded as a kind of 
 Schwinger-Dyson equation. In particular, this identity reduces in the semiclassical  limit to the triviality of the product of local monodromies $\Omega_1\Omega_2\Omega_3=1$ around the vertex operators,  which played a crucial role in   providing all important global information on the three-point 
 function in the strong coupling regime \arXiv{1312.3727}{1312.3727}. This structure may provide a key to the understanding of the notion of ``integrability" beyond the spectral level. 
\newpage
\baselineskip 3.3ex
\topmargin -1.3cm
\oddsidemargin -0.4cm \evensidemargin 0cm
\textwidth=16cm 
\renewcommand{\thefootnote}{\arabic{footnote}}
\thispagestyle{empty}
\enlargethispage{2\baselineskip}
\renewcommand{\contentsname}{{\small \flushleft{Contents}}}
{\footnotesize \tableofcontents}
\section{Introduction\label{sec:intro}}
Among the multitude of  quantities investigated for the understanding of the 
 AdS/CFT duality \cite{AdSCFT1, AdSCFT2, AdSCFT3}, the three-point functions of the gauge-invariant 
 composite operators in the $\mathcal{N}=4$ super Yang-Mills theory in the weak 
 and the strong coupling regimes are perhaps the most basic 
 objects that directly probe the correspondence of dynamical interactions 
in the prototypical setting.  In particular, in the simplest sector called 
 the ``SU(2) sector"\cite{MZ,KMMZ}, there have been substantial progress in both the 
 weak and the strong coupling regimes in the past few years.  

For the weak coupling perturbative computation\footnote{For earlier pioneering investigations, see \cite{OT, RV, ADGN}.}, a systematic procedure called 
 ``tailoring" has been developed \cite{Tailoring1, Tailoring2, Tailoring3, quantum3pt, Tailoring4}, and with a useful technical improvement \cite{Foda}, a special class of 
 three-point functions for non-BPS operators have been expressed explicitly in terms of Slavnov determinants \cite{Slavnov}. Furthermore,  the  semi-classical 
 limit of such three point functions with large charges were successfully evaluated in a remarkably  compact form \cite{Tailoring3, Kostov1, Kostov2, Fixing}. 

On the other hand, the strong coupling computation was performed using  the string theory in $AdS_3\times S^3$ spacetime \cite{KK3}, with the vertex operators possessing the same global quantum numbers  as the operators in 
the  ``SU(2) sector"  considered at weak coupling.  Since the canonical  quantization of the string  in such a curved space is not available at present, 
the saddle point approximation was used, which is valid for the case of vertex 
 operators carrying  large charges. Although the precise form of the vertex 
 operators nor the exact saddle  point configuration  were not known, the 
judicious use of classical integrability,  with  a  certain natural  assumption, 
was  powerful enough  to produce explicit answers for the desired three-point 
 functions.  Surprisingly, even before taking any limits, the results 
exhibited  structures rather similar to those at weak coupling. On the other hand, upon taking  the so-called Frolov-Tseytlin limit, in which the strong and the weak coupling  results were expected to agree, small discrepancies were 
observed, the understanding of which is left as a future problem. 

Evidently, besides making the comparison of the results,  the principal goal of these investigations is to uncover common concepts  and structures threading the both sides of the duality and understand how they are realized to make the
 duality work. For this purpose, it is desirable to be able to treat the both sides 
 in as much the same way as possible and try to extract the key principle. 
In this article, we shall present two new significant results in the weak coupling 
 analysis  for such a purpose, 
which are actually  hinted by the strong coupling investigation of \cite{JW,KK1,KK2,KK3}. 
Let us now briefly describe them one by one. 

The first result concerns the computation of the three-point functions much more
 general than those  treated so far in the existing literature.  
As is customary,  let  $\Phi_i$ $(i=1,2,3,4)$ be the four of the six adjoint scalar fields 
forming the SU(2) sector and  denote their complex combinations as 
\begin{align}
\begin{aligned}
Z&=\Phi_1+i\Phi_2\comma \qquad   \Zbar = \Phi_1-i\Phi_2 \comma \\
 X&=\Phi_3+i\Phi_4\comma \qquad \Xbar=\Phi_3-i\Phi_4 \period
\end{aligned}
\end{align}
In the systematic investigation initiated  in   \cite{Tailoring1}, two of  the three operators  interpretable as XXX${}_{1/2}$ spin chains were taken to be built  upon 
 the pseudo-vacuum $\trace(Z^\ell)$,  and the  remaining one was built upon $\trace(\Zbar^\ell)$.  As long as one identifies $Z$ and $\Zbar$ as  ``ground state" up-spins  and $X$ and  $\Xbar$ as down-spins  representing  the excitations, such a choice of operators 
were essentially unique in order to produce  non-extremal correlators. 

In the work of \cite{KK3}, however, a detailed analysis has been made of 
  the operators built upon  more general ``vacuum" states 
where an arbitrary  linear combination of  $\Phi_i$ is regarded as 
  the ``up-spin".  This study revealed that the natural way to characterize 
 the general operators so constructed is by a pair of two-component vectors $n$ and $\tilde{n}$, 
termed ``polarization spinors", associated  to each of the SU(2) factors 
 of the global symmetry group  SO(4) $\cong$ SU(2)$_L\times$SU(2)$_R$. 
By applying this characterization to  the string vertex operators, three-point 
functions of operators carrying general polarization spinors were computed at strong coupling. 

Since the analysis of the general operators mentioned above was inspired  in the spin-chain picture of  the operators, one would expect that similar generalization can and should be done at weak coupling. Clearly this would be important in the comparison with the strong coupling results. Unfortunately, however, there are apparent problems to overcome. One is that when the three operators 
 are built on different ``rotated vacua", it is non-trivial to perform the 
Wick contractions keeping the spin-chain interpretation intact. Another difficulty  is that,  for the general configurations under consideration,  $\langle \text{off-shell}|\text{off-shell}\rangle$ inner products  produced through  the usual tailoring procedure cannot in general be converted  into $\langle \text{on-shell}|\text{off-shell}\rangle$ form by the known trick \cite{Foda}. This hampers the expressions in terms of tractable determinants. 

As will be explained fully in sections \ref{sec:double} and \ref{sec:Wick}, these problems will be neatly solved by (i) the ``double spin-chain" formulation of the conventional  spin-chain and  (ii) the novel interpretation of the Wick contraction as skew-symmetric singlet  paring acting  on the double spin-chain Hilbert space. 
These ideas allow us to characterize  the general operators by a pair of 
polarization spinors and moreover naturally  factorize the three-point functions into the product of \SUL and \SUR  factors, just as it happened for the wave function  part of the strong coupling computation\cite{KK3}. 
The most important advantage, however, is the fact that under the new singlet  pairing interpretation, the Wick contraction procedure produces only the matrix elements of the $B(u_i)$ components of the monodromy matrix,  without the appearance of $C(u_i)$ components.  Therefore the building blocks of 
 the three-point functions  take the form of  the so-called partial domain 
wall partition function (pDWPF) \cite{Korepin, Izergin, FW,Kostov1,Kostov2,KM} and immediately possess determinant expressions. In particular, for certain class of correlators the expression  in terms of the sum of  pDWPF's collapses into a single term and 
yields  a remarkably simple result. 

Now let us move on to the second new result, which  again is motivated 
 by the structure of the strong coupling computations \cite{JW, KK1, KK2, KK3}. One of the crucial difficulties  in the strong coupling computation is that 
 one does not know the exact  three-pronged saddle point solution with which to evaluate the three-point function. In the framework of the classical integrable system, the most important available information is the form 
 of the solution of the auxiliary linear problem (ALP) in the vicinity of the vertex operator insertion point $z_i$, which can be approximated\footnote{
Actually, as far as the evaluation of the wave function for the three-point 
 function is concerned,  the slight deviation from the two-point function near the puncture contains  a crucial information\cite{KK3}.}
 by  the saddle point configuration for the two-point solution. Differently put, 
the local monodromy operator $\Omega_i$ and  its linearly independent eigenfunctions $i_\pm$ of ALP  around $z_i$ are the only available secure yet local data.  It is clear that in addition one definitely needs
some global information to capture the properties of the three-point function. 
As was demonstrated  in the previous works \cite{JW, KK1, KK2, KK3},  
such a global information  was provided by 
 the triviality of the total monodromy, namely $\Omega_1\Omega_2\Omega_3=1$. This seemingly weak constraint turned out to be  surprisingly powerful and played a key role in  computing the Wronskians of the eigenfunctions $\langle i_+ , j_+ \rangle$,  etc.  with which the three-point functions are constructed. 

This experience strongly suggests that one should formulate a similar monodromy relations for the three-point functions at weak coupling as well.  The corresponding quantities are  the three-point functions with three  local monodromy operators inserted. As will be explained in section \ref{sec:mono}, non-trivial relations, which contain certain constant shifts of the spectral parameter, can be obtained through the use of the so-called ``unitarity" and ``crossing'' relations for the Lax operator. 
Generically such monodromy identity relates three-point functions 
 composed of different operators and hence may be regarded as a kind of 
Schwinger-Dyson equation.  As a simple application, one can obtain the 
counterpart of the total trivial monodromy relation $\Omega_1\Omega_2\Omega_3=1$ in the semi-classical limit of the large spectral parameter, where the constant 
 shifts can be ignored.  Just as in the case of strong coupling, such a relation 
 provides vital information in the computation of the three-point functions, 
the details of which will be fully described  in a separate communication\cite{KKN3}. 
In any event, this structure may provide a key to the understanding of the 
notion of ``integrability" beyond the spectral level, 
especially if it can be generalized to higher loop correlators. 

The organization of the rest of the  article is as follows: In section \ref{sec:double}, 
 we will begin by explaining the double spin-chain formalism for the 
SU(2) sector and introduce the general rotated vacua and construct the 
non-BPS operators built upon such vacua. Then in section \ref{sec:Wick}, we will 
 formulate the new group-theoretical view of the Wick contractions of constituent fields and the composite operators made out of them, which is natural for the  double spin-chain formulation. With theses preparations, we will describe in section \ref{sec:threepnt} how one can compute the three-point functions which are much more general than the ones considered so far in the literature. The advantage of our new formalism becomes apparent in this computation in that the correlators factorize into the \SUL\!- and the \SUR\!- pieces and will be naturally expressed in terms of the determinants which describe the partial domain wall partition functions. The new global monodromy relations 
 for the three-point functions will be derived in section \ref{sec:mono}. In the double spin-chain formalism, this relation will also enjoy the factorized properties. 
Finally in section \ref{sec:discuss}, we will discuss future directions and briefly comment on 
 a direct computation of the semi-classical three-point functions 
without the use of the determinant formulas, being prepared as a separate 
treatise \cite{KKN3}. Two short appendices are provided to explain the kinematical dependence of the three-point functions and the general form of the monodromy relation.

Note:  We acknowledge that a part of the subjects discussed in this paper is
 also investigated independently in the recent paper by Y.~Jiang, I.~Kostov, A.~Petrovskii and D.~Serban \cite{spinvertex}.

\section{Double spin-chain formalism for the SU(2) sector\label{sec:double}} 
As described in the introduction, one of the two major aims of this 
paper is to develop a scheme  in which the three-point functions 
of a more general class in the SU(2) sector can be computed systematically. 
This is of value since such a computation has already been done 
 in the strong coupling regime\cite{KK3} and it is important to be able to make a comparison of their general structures. In this section, we shall explain the 
basic idea of this formalism, to be called the ``double spin-chain formalism". 
\subsection{Scalar fields as tensor products of two spins\label{subsec:scalar}}
Let us begin with the description of a new way of mapping each of the  basic fields  $Z, \Zbar, X, \Xbar$ of the ``SU(2)" sector to a  tensor product  of 
two spin-chain states.  In the previous approach\cite{Tailoring1}, one makes the identifications of  the basic up- and down- spin pair as 
 $(Z, X) \mapsto (\upket, \downket)$, $(Z, \Xbar) \mapsto (\upket, \downket)$ and $(\Zbar, \Xbar) \mapsto (\upket, \downket)$\footnote{In the ``tailoring" formulation \cite{Tailoring1}, the pair $(\Zbar, X)$ is not needed for the construction of three distinct spin-chains 
making up  the three-point functions.}. Thus, although the content of these three pairs are obviously  different, the spin chains composed of them are regarded as the {\it same type} of SU(2) spin chain.  This somewhat redundant 
characterization of the constituents of the spin chains makes it difficult 
 to construct  the correlators  of three operators forming 
 spin chains where their relevant SU(2) groups are embedded in  more general ways  in the total symmetry group SO(4). 

A natural and simple solution to this problem is to make use of the fact that 
 the basic fields $(Z, \Zbar, X, \Xbar)$ carry distinct charges with respect to 
SU(2)$_L\times$SU(2)$_R$ $(\cong $SO(4)$)$. This is best expressed by 
 assembling them into the $2\times 2$ matrix 
\beq{
\Phi_{a\tilde{a}}\equiv\pmatrix{cc}{Z&X\\-\bar{X}&\bar{Z}}_{a\tilde{a}} \comma \label{eq-1}
}
which transforms as 
\begin{align}
\Phi \rightarrow U_L \Phi U_R \comma  \label{LRtransf}
\end{align}
where $U_L \in$ SU(2)$_L$ and $U_R \in$ SU(2)$_R$. This means that  these fields carry left and the right charges $(L, R)$ of the form 
\beq{
\begin{aligned}
&Z: \,(+1/2, +1/2)\comma\quad  &X: \,(+1/2, -1/2)\comma\\
&\bar{Z}: \,(-1/2, -1/2)\comma \quad &-\bar{X}: \,(-1/2, +1/2)\period
 \end{aligned}
 }
Thus, from the representation-theoretic point of view, it is natural to map each of these fields  to a tensor product of two spin-states in the following way:
 \beq{
 \begin{aligned}
 &Z\mapsto \ket{\upspin}_L\otimes \ket{\upspin}_R\comma &X\mapsto \ket{\upspin}_L\otimes \ket{\downspin}_R\comma\\
 &\bar{Z}\mapsto \ket{\downspin}_L\otimes \ket{\downspin}_R\comma &-\bar{X}\mapsto \ket{\downspin}_L\otimes \ket{\upspin}_R\comma
 \end{aligned}\label{eq-2}
 }
This evidently leads to the double spin-chain formalism, which will be much more versatile than the conventional single spin-chain treatment. 
As an example, consider a general linear combination of the four fields, which can be written as 
\begin{align}
\P \cdot \Phi \equiv \sum_{a, \atil} \P^{a\atil} \Phi_{a \atil} \comma 
\end{align}
where $\P^{a\atil}$ is a $2\times 2$ matrix. Then, clearly this quantity  maps to the double spin-chain state as 
 \beq{\P\cdot\Phi\mapsto
 \P^{1\tilde{1}} \ket{\upspin}_L\otimes \ket{\upspin}_R +\P^{1\tilde{2}} \ket{\upspin}_L\otimes \ket{\downspin}_R + \P^{2\tilde{1}}\ket{\downspin}_L\otimes \ket{\upspin}_R +\P^{2\tilde{2}}\ket{\downspin}_L\otimes \ket{\downspin}_R\period
 }
\subsection{General rotated vacua\label{subsec:BPS}}
Next let us turn to the construction and the description of the general spin-chains. To do this, we  must first prepare a general vacuum state upon which 
 the SU(2) magnon excitations are created. The most transparent way 
 to construct such a general vacuum state is to make an arbitrary  $\SULR$ 
transformation to  the conventional BPS vacuum state $\trace(Z^\ell)$, where
 $\ell$ is the length of the spin-chain. Under the transformation \eqref{LRtransf}, $Z$ itself turns into 
\begin{align}
Z &= (\Phi)_{11} \longrightarrow (U_L \Phi U_R)_{11} 
 = (U_L)_1{}^a \Phi_{a\atil} (U_R)^{\atil}{}_1
\end{align}
Comparing this with the general linear combination $\P^{a\atil} \Phi_{a\atil}$, we learn that $\P^{a\atil}$ can be written as a product 
\begin{align}
&\P^{a\atil} = \n^a\tilde{\n}^{\atil}   \comma \label{Pnn}\\
&\n^a =  (U_L)_1{}^a\comma \qquad \tilde{\n}^\atil = (U_R)^{\atil}{}_1\nn
\end{align}
Hereafter, we use the notations where  the indices $a$ and $\atil$
are lowered and raised  by the $\ep$ tensors $\ep_{ab}, \ep_{\atil\btil}, \ep^{a\atil},  \ep^{a\atil}$,  with 
 the convention $\ep_{12} =1, \ep^{12} =1$. This means $\ep_{ab}\ep^{bc} = -\delta_a^c$ and $\ep_{ab} \ep^{ab} =2$, etc.  For instance, 
$\P_{a\atil}$ is defined as $\P_{a\atil}=\ep_{ab}\ep_{\atil\btil} \P^{b\btil}$. Then it is easy to see that $\P^{a\atil}$ is nilpotent in the sense 
 that  $\P^{a\atil}\P_{a\atil} = \n^a\tilde{\n}^\atil \ep_{ab}\ep_{\atil\btil} \n^b \tilde{\n}^\btil=0$.

Now because of the structure \eqref{Pnn}, the combination $\P \cdot \Phi$  is mapped to the spin state
\begin{align}
\P\cdot \Phi \mapsto \ket{\n}_L \otimes \ket{\tilde{\n}}_R \comma 
\end{align}
where 
\beq{
&\ket{\n}_{L}\equiv \n^1\ket{\upspin}_L +\n^2\ket{\downspin}_L \comma\qquad
\ket{\tilde{\n}}_R\equiv\tilde{\n}^1\ket{\upspin}_R +\tilde{\n}^2\ket{\downspin}_R\period \label{defketn}
}
This makes it clear that the two dimensional vectors $\n^a$ and $\tilde{\n}^\atil$ characterize the scalar fields completely. Such vectors were introduced in \cite{KK3} and  were termed ``polarization spinors"\fn{Note that the notation for the polarizations is slightly different from the one in \cite{KK3}: In \cite{KK3}, we denoted the SU(2)$_L$ polarization spinor by $\tilde{n}$ and SU(2)$_R$ polarization spinor by $n$.}. 

It is now easy to see  that the rotated BPS vacuum 
\beq{
\trace\left(({\rm P}\cdot \Phi)^{\ell}\right) \label{eq-4}
}
 is mapped to the spin-chain state of the form 
\beq{
\trace\left(({\rm P}\cdot \Phi)^{\ell}\right) \mapsto \ket{\n^{\ell}}_L\otimes\ket{\tilde{\n}^{\ell}}_R\comma\label{eq-5}
 } 
where $\ket{\n^{\ell}}_L$ and $\ket{\tilde{\n}^{\ell}}_R$ are 
 given by 
 \beq{
 \ket{\n^{\ell}}_L = \underbrace{\ket{\n}_L\otimes\cdots\otimes \ket{\n}_L}_{\ell}\comma\quad  \ket{\tilde{\n}^{\ell}}_R = \underbrace{\ket{\tilde{\n}}_R\otimes\cdots\otimes \ket{\tilde{\n}}_R}_{\ell}\period
 }
 For later convenience, we impose the following normalization conditions  on the polarization spinors:
 \beq{
 \n^{a}\overline{\n}_a=1\comma \quad \tilde{\n}^{\tilde{a}}\overline{\tilde{\n}}_{\tilde{a}}=1\comma\label{eq-6}
  } 
 where the ``conjugate spinors'' $\overline{\n}$ and $\overline{\tilde{\n}}$ are defined by
 \beq{
 \overline{\n}_{a}\equiv \left( \n^a\right)^{\ast} \comma \quad \overline{\tilde{\n}}_{\tilde{a}}\equiv \left( \tilde{\n}^{\tilde{a}}\right)^{\ast}\comma
 }
 The condition \eqref{eq-6} determines the normalization of the operator \eqref{eq-4} up to a phase. The phases of the operators only affect the overall phase of the structure constant, which we will not discuss in this paper.
\subsection{Non-BPS operators as excitations on rotated vacua\label{subsec:nonBPS}}
We will now express non-BPS operators as excited states on 
 the general rotated vacua constructed in the previous subsection. 

The strategy is straightforward.  We will first consider the 
 excited states built upon the conventional vacuum $\ket{\uparrow^\ell}$ 
in both the left and the right sectors by the algebraic Bethe ansatz procedure. 
Explicitly, the states obtained are 
\beq{
\ket{\bm{u};\uparrow^{\ell}}_L = B(u_1)\cdots B(u_M)\ket{\upspin^{\ell}}_L\comma \qquad \ket{\tilde{\bm{u}};\uparrow^{\ell}}_R = B(\tilde{u}_1)\cdots B(\tilde{u}_{\tilde{M}})\ket{\upspin^{\ell}}_R\comma\label{aba}
}
where the sets of rapidities $\bm{u}$ and $\tilde{\bm{u}}$ are assumed to satisfy the Bethe equation. As is customary, the magnon creation operator $B(u)$ is defined through the monodromy matrix as
\beq{
\begin{aligned}
&\Omega(u)\equiv{\rm L}_1 (u-\theta_1){\rm L}_2 (u-\theta_2)\cdots {\rm L}_{\ell} (u-\theta_{\ell})= \pmatrix{cc}{A(u)&B(u)\\C(u)&D(u)}\comma\\
&\qquad {\rm L}_k(u)=\pmatrix{cc}{u+iS_3^{k}&iS_{-}^{k}\\iS_{+}^{k}&u-iS_3^{k}}\comma
\end{aligned}
}
where $S_{\ast}^k$ denotes the SU(2) spin operator acting on the $k$-th site of the spin chain and ${\rm L}_k(u)$ is the Lax operator associated to site $k$. The extra parameters $\theta$'s  introduced here are called the inhomogeneities. To compute the tree-level correlation functions, we do not need such parameters and they should be simply set to zero. However, as discussed in \cite{quantum3pt,Tailoring4,Serban,Fixing}, the inhomogeneities are known to be useful for discussing the loop corrections to the three-point functions. Therefore, we will keep them in the following discussions.

 Now in order to obtain the state which can be 
 interpreted as an SU(2) spin-chain, we may  excite either the left sector 
 or the right sector, but not both. If we excite both, such a state cannot be 
obtained by any embedding of SU(2) in SO(4).  Therefore, we have
the following two types of excited states, which we call type I and type II:
\beq{
\text{Type I}: \ket{\bm{u};\uparrow^{\ell}}_L \otimes \ket{\upspin^{\ell}}_R\comma \qquad \text{Type II}: \ket{\upspin^{\ell}}_L\otimes \ket{\tilde{\bm{u}};\uparrow^{\ell}}_R\period  \label{eq-7}
}
It is important to note that they cannot be related by an  SO(4) rotation since 
 there is no transformation within SO(4) which interchanges \SUL and \SUR.

Once we have these basic states, we can now rotate them by an arbitrary  $\SULR$  transformation to produce general excited states. A very useful way to 
 parametrize  the SU(2)$_L$ and SU(2)$_R$ transformations is as follows. 
As shown previously the polarization spinors characterize the rotated fields precisely. Therefore one can specify, for example, an element  $\g_{\n} \in $
SU(2)$_L$ by the equation 
\begin{align}
\g_{\n} \ket{\!\uparrow}_L = \ket{\n}_L \comma 
\end{align}
 up to a phase coming from the U(1) rotation $h$  which leaves $\ket{\upspin}$ invariant. Since  we shall ignore such a phase in this work, 
what is relevant is actually the parametrization of the coset SU(2)$/$U(1),
the element of which will be denoted by  $g_{\n}$, where 
\begin{align}
\g_{\n}=g_{\n} h\comma \qquad \g_{\n} \in {\rm SU(2)} \comma \quad g_{\n}\in {\rm SU(2)}/{\rm U(1)}
\comma \quad h \in {\rm U(1)} \period
\end{align}

Among the various parametrizations of SU(2)$/$U(1), the one which will  be most useful is the so-called the coherent state parametrization.  In the spin $1/2$ highest weight representation we are adopting, the useful expression  for the 
 coset element $g_{\n}$ is 
obtained by the SU(2) Baker-Campbell-Hausdorff formula in the form\cite{Zhang}
\begin{align}
g_{\n} = e^{-\bar{\zeta} S_+ +\zeta S_-} &= e^{z  S_-} e^{-\ln (1+|z|^2) S_3} e^{-\zbar S_+} \comma \label{subch1}\\
&= e^{-\zbar S_+} e^{\ln (1+|z|^2) S_3} e^{z S_-}  \label{subch2}  
\end{align}
where $z = (\zeta /|\zeta|) \tan |\zeta|$ and $S_i$'s are the generators of 
 the global SU(2), with the convention $S_\pm \equiv S_1\pm i S_2$. 
Since $\ket{\!\uparrow}_L$ corresponds to $\n^a =(1,0)^t$, applying 
\eqref{subch1} we get
\begin{align}
\n^a &= g_{\n} \vecii{1}{0} ^a= {1\over \sqrt{1+|z|^2}} \vecii{1}{z} ^a
\period 
\end{align}
Similarly, coset elements corresponding to $\ket{\!\uparrow}_R, \ket{\!\downarrow}_L, \ket{\!\downarrow}_R$ are  characterized by 
\beq{
 g_{\n}\ket{\downspin}_L= \ket{\overline{\n}}_R\comma
\qquad 
\tilde{g}_{\tilde{\n}}\ket{\upspin}_L = \ket{\tilde{\n}}_R\comma \qquad \tilde{g}_{\tilde{\n}}\ket{\downspin}_R= \ket{\overline{\tilde{\n}}}_R\comma \label{eq-9}
 }
and the corresponding polarization spinors can be computed similarly,  using \eqref{subch1} or \eqref{subch2} where appropriate, as\footnote{We redisplay the result for $\n^a$ as well for convenience.}
\begin{align}
\n^a &=  {1\over \sqrt{1+|z|^2}} \vecii{1}{z} ^a 
\comma \qquad \bar{\n}^a =   {1\over \sqrt{1+|z|^2}} \vecii{-\zbar}{1} ^a  \comma \label{nanbara1}\\
\tilde{\n}^\atil &=  {1\over \sqrt{1+|\tilde{z}|^2}} \vecii{1}{\tilde{z}} ^a 
\comma \qquad \bar{\tilde{\n}}^\atil =   {1\over \sqrt{1+|\tilde{z}|^2}} \vecii{-\bar{\tilde{z}}}{1} ^\atil \period \label{nanbara2}
\end{align}
With this preparation, it is now straightforward to write down the general  excited states of type I and II built upon the rotated vacuum $\ket{\n^{\ell}}\otimes \ket{\tilde{\n}^{\ell}}$ as 
\beq{
&\text{Type I}: \ket{\bm{u};\n^{\ell}}_L \otimes  \ket{\tilde{\n}^{\ell}}_R\comma \qquad
\text{Type II}:\ket{\n^{\ell}}_L\otimes \ket{\tilde{\bm{u}};\tilde{\n}^{\ell}}_R\comma\label{eq-11}
}
where $\ket{\bm{u};\n^{\ell}}_L$ and $\ket{\tilde{\bm{u}};\tilde{\n}^{\ell}}_R$ are obtained by the SU(2)$_L$ and SU(2)$_R$ rotations discussed above:
\beq{
\ket{\bm{u};\n^{\ell}}_L\equiv g_{\n}\ket{\bm{u};\upspin^{\ell}}_L\comma \quad \ket{\tilde{\bm{u}};\tilde{\n}^{\ell}}_R\equiv\tilde{g}_{\tilde{\n}}\ket{\tilde{\bm{u}};\upspin^{\ell}}_R\period\label{eq-12}
}

Now it is well-known that, when the rapidities $\bm{u}$ and $\tilde{\bm{u}}$ are all finite, the on-shell Bethe states constructed upon the up-spin vacuum \eqref{aba} satisfy the highest weight condition 
\beq{
S_{+}\ket{\bm{u};\uparrow^{\ell}}_L=0\comma \qquad S_{+}\ket{\tilde{\bm{u}};\uparrow^{\ell}}_R=0\period\label{hw}
}
Upon such states, the actions of $g_{\n}$ and $\tilde{g}_{\tilde{\n}}$ 
simplify because the last factor in \eqref{subch1} becomes unity. As a result, 
we obtain the following expressions\fn{The idea to characterize the rotated state in a similar way was proposed previously in \cite{Kostov_IGST}.}:
\beq{
\begin{aligned}
&\ket{\bm{u};\n^{\ell}}_L = \left( \frac{1}{1+|z|^2}\right)^{\ell/2 -M}\,e^{zS_{-}}\ket{\bm{u};\upspin^{\ell}}_L\comma\\
&\ket{\tilde{\bm{u}};\tilde{\n}^{\ell}}_R=\left( \frac{1}{1+|\tilde{z}|^2}\right)^{\ell/2 -\tilde{M}}e^{\tilde{z}S_{-}}\ket{\tilde{\bm{u}};\upspin^{\ell}}_R\period
\end{aligned}\label{eq-13}
}
We shall see that the representations  \eqref{eq-13} will be quite useful when we evaluate the three-point functions in section \ref{sec:threepnt}. 
\section{Wick contraction as skew-symmetric singlet pairing \label{sec:Wick}}
Having prepared the operators interpretable as general classes  of spin-chains built upon rotated vacua, we now discuss how to perform the Wick contractions of such objects in an efficient manner based on a  group-theoretical 
point of view.  
\subsection{Wick contraction for general constituent fields\label{subsec:Wickfields}}
To begin, let us  discuss the Wick contraction of the constituent fields. 
At the tree level, the contraction rules for the basic complex scalar fields are 
given by 
\beq{
\ewick{Z}{Z} = 0\comma \quad \ewick{Z}{X}=0\comma \quad \ewick{Z}{\bar{X}}=0 \comma \quad \ewick{Z}{\bar{Z}} =1\comma  \quad \text{etc}\period
 }
It will be most useful to regard these rules as those for the 
  elements of the matrix $\Phi_{a\atil}$ given in \eqref{eq-1}.  It is easy to 
check that the above rules are neatly summarized as 
\begin{align}
\ewick{\Phi_{a\atil}}{\Phi_{b\btil}} &= \ep_{ab} \ep_{\atil\btil} \period
\label{basicwick}
\end{align}
Now recall that the general linear combination of these fields can be  
written as 
\begin{align}
\P^{a\atil}\Phi_{a\atil} = \n^a  \tilde{\n}^\atil \Phi_{a\atil} \comma 
\label{eq-14}
\end{align}
where we 
 used the factorized expression of $\P^{a\tilde{a}}$ in terms of the polarization 
 spinors \eqref{Pnn}. Then, using  \eqref{basicwick} and \eqref{eq-14}, the contraction of two general  combinations 
denoted  as  $F_1=\P_1 \cdot \Phi$ and $F_2=\P_2 \cdot \Phi$ 
can be  immediately computed as 
 \beq{
 \ewick{F_1}{F_2}=
 \left(\n_1^{a}\n_{2}{}_a\right)\left(\tilde{\n}_1^{\tilde{a}}\tilde{\n}_{2}{}_{\tilde{a}}\right)\period\label{eq-15}
 }
This formula reveals  that in terms of the polarization spinors the Wick contraction is nothing but the operation of forming singlets in both the SU(2)$_L$ and the SU(2)$_R$ sectors. 

\begin{figure}[t]
  \begin{center}
   \picture{clip,height=2.5cm}{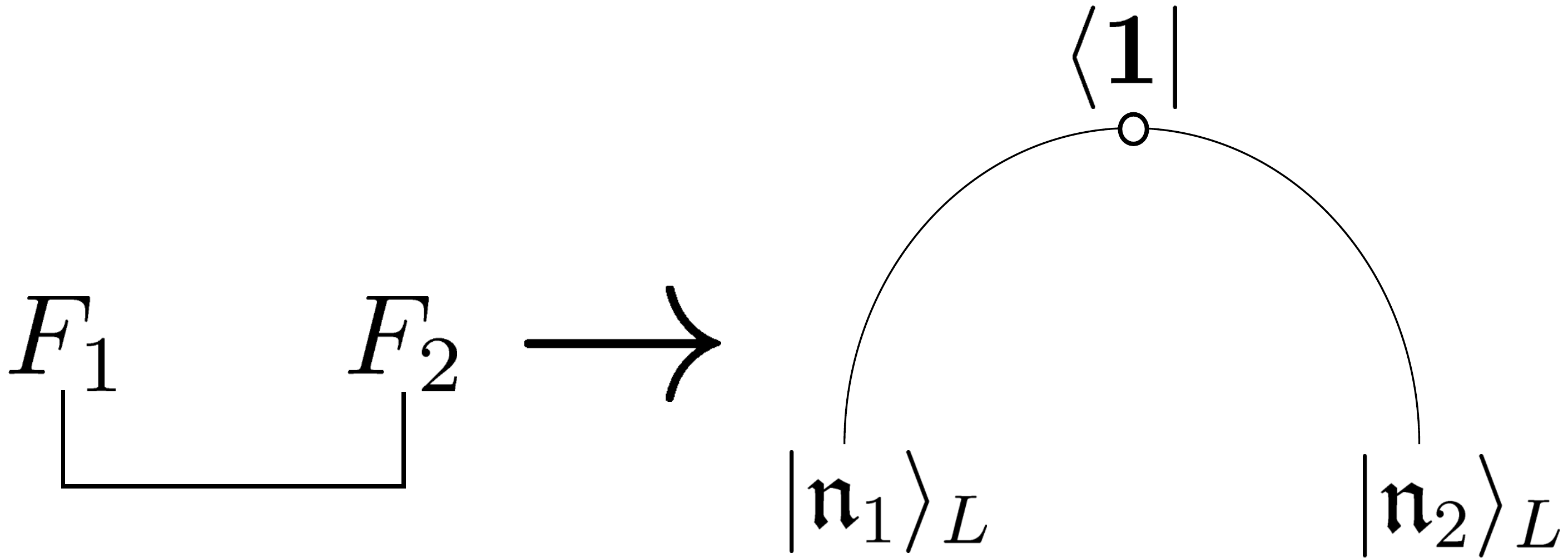}
  \end{center}
\caption{The wick contraction and the singlet pairing \eqref{eq-16}. The white blob denotes the singlet state $\bra{\s}$. Here we only depicted SU(2)$_L$ sector.} 
\label{pairing}
\end{figure}

We now would like to transplant  this structure in the spin-chain language. 
For this purpose, it is convenient to write the up and the down spin state 
collectively as $\ket{a}$ with the definition\fn{Of course we do this for both the left and the right sectors. Here for simplicity we suppress  the subscripts $L$ and $R$,  as the structure is common.}
\begin{align}
\begin{aligned}
\ket{a} :&\quad \ket{1} \equiv \ket{\uparrow} \comma 
\quad \ket{2} \equiv \ket{\downarrow} \comma  \\
& \langle a \ket{b} = \delta_{ab} \period
\end{aligned}\label{defspinstate}
\end{align}
Then, from the definition of $\ket{\n}$ given  in \eqref{defketn} 
 we have 
\begin{align}
\ket{\n} &= \n^a \ket{a} \comma \qquad \n^a = \langle a \ket{\n} 
\period 
\end{align}
Let us now introduce the singlet projection operator $\bra{\s}$ 
 in the following way:
\begin{align}
\bra{\s} &\equiv \ep_{ab} \bra{a} \otimes \bra{b} \period \label{singproj}
\end{align}
When acted on the state of the form $\ket{\n_1} \otimes \ket{\n_2}$, it projects out  the singlet in the manner
\begin{align}
\bra{\s} \left( \ket{\n_1} \otimes \ket{\n_2} \right) 
 =\ep_{ab} \langle a \ket{\n_1} \langle b \ket{\n_2} = \ep_{ab} \n_1^a
\n_2^b = \n_1^a \n_{2a} \period 
\end{align}
Therefore the contraction $\ewick{F_1}{F_2}$ given in \eqref{eq-15} 
 is reproduced as 
\begin{align}
\ewick{F_1}{F_2} &= \bra{\s} \left( \ket{\n_1}_L \otimes \ket{\n_2}_L \right)   \bra{\s} \left( \ket{\tilde{\n}_1}_R \otimes \ket{\tilde{\n}_2}_R \right)  
\label{eq-16}
\end{align}
This relation is expressed pictorially in \figref{pairing}.
Note that each factor on the right hand side of \eqref{eq-16} is anti-symmetric under the interchange of two spin states, unlike the ordinary inner product used in the previous works \cite{Tailoring1,Tailoring2,Tailoring3,Tailoring4,Foda,Fixing}. 

It should be remarked that the appearance of the singlet state in the expression \eqref{eq-16} is quite natural from a physical point of view: Every Feynman diagram, including the ones with vertices, can be viewed, from an  appropriate direction, as a virtual process in which the fields annihilate into the vacuum. Since the vacuum is not charged under any symmetry, it belongs to the singlet representation for all the symmetry groups. Thus, different Feynman diagrams account for different ways of producing the singlet representation starting from a given field-configuration. The simplest way to achieve this is to take a pair of fields and project it to the singlet representation, which is exactly what \eqref{eq-16} does. This argument suggests that the singlet state will play an important role also in other sectors\fn{Although our motivation was to
 provide a new interpretation for the Wick contraction of the fields
 forming a spin chain, a very similar idea of invariant pairing was introduced in a different
 context, namely the mapping from CFT${}_4$ to TFT${}_2$ in \cite{KR}. 
 Thieir description is likely to be quite useful for the
 construction of three-point functions for the non-compact sectors of
 the PSU(2,2$|$4) spin chain.} and at  higher-loop order, although the expression will certainly be more complicated than \eqref{eq-16}.
\subsection{Wick contraction for two composite operators\label{subsec:Wickop}}
Let us next express the Wick contraction between two composite operators $\mathcal{O}_1$ and $\mathcal{O}_2$ using the skew-symmetric  inner product defined above. In what follows, we denote the spin-chain states corresponding  to the operators $\mathcal{O}_1$ and $\mathcal{O}_2$ abstractly as\fn{$\mathcal{O}_1$ and $\mathcal{O}_2$ can be either of type I or type II in \eqref{eq-11}.}
\beq{
\mathcal{O}_1\mapsto \ket{\mathcal{O}_1}_L\otimes \ket{\tilde{\mathcal{O}}_1}_R\comma \qquad \mathcal{O}_2\mapsto \ket{\mathcal{O}_2}_L\otimes \ket{\tilde{\mathcal{O}}_2}_R\period
}
As we are working in the large $N_c$ limit, the unsuppressed  Wick contractions between two composite operators are of a special type, an example of 
 which is given by 
\beq{
\tr\big(\cdots \setlength{\underbracketheight}{9pt}\cunderbracket{$X$}{\setlength{\underbracketheight}{3pt}\cUnderbracket{$Z$}{$\big)\qquad\tr\big($}{$ \bar{Z}$}}{$\bar{X}$}\cdots \big)\period
}
The structure  should be clear:  the allowed contractions are between the rightmost field in $\mathcal{O}_1$ with the leftmost field in $\mathcal{O}_2$ 
and so on, as indicated. Obviously the two spin chains must be of the 
same length to be non-vanishing under the contractions. 

\begin{figure}[t]
  \begin{center}
   \picture{clip,height=3cm}{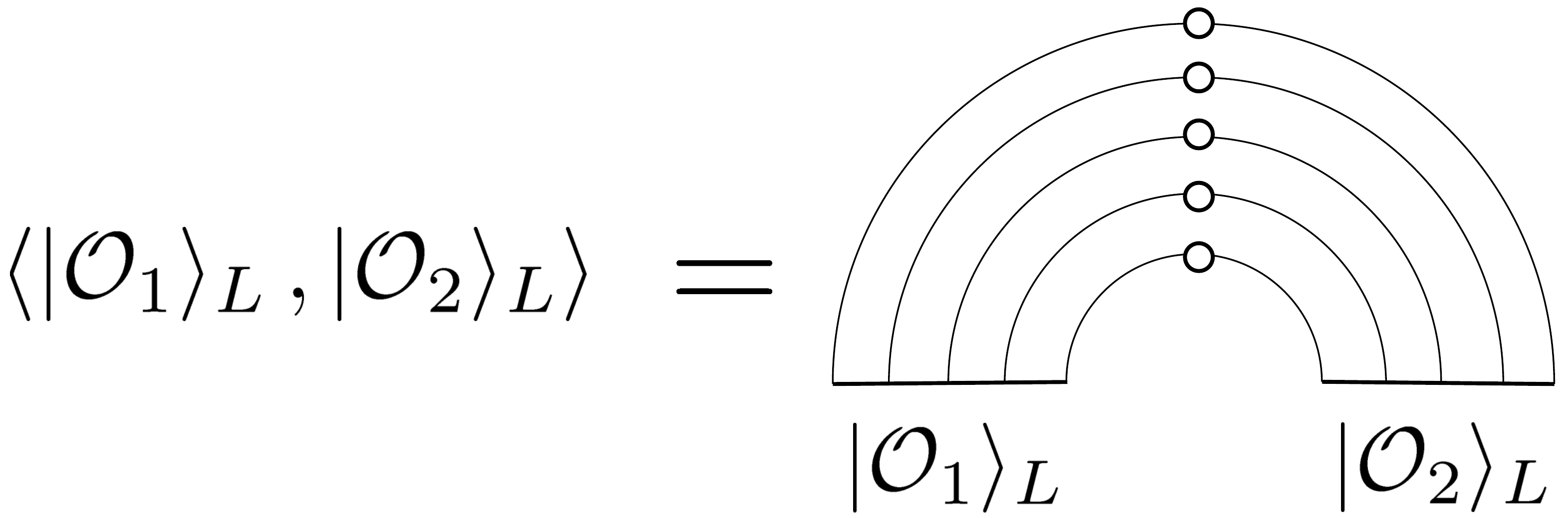}
  \end{center}
\caption{A pictorial definition of the skew-symmetric inner product for two spin-chain states: We first compute an overlap between the singlet state $\bra{\s}$ and a tensor product of two spins connected to a single blob ($\circ$), and then, take a product of such overlaps. Here we only described the SU(2)$_L$ chain. The definition for the SU(2)$_R$ chain is basically the same.} 
\label{2pt_comp}
\end{figure}

This type of contraction rule is expressed in the spin-chain language by using 
 the following skew-symmetric inner product between two states of the same length, $\ket{\Psi_1}$ and $\ket{\Psi_2}$:
\beq{
\big< \ket{\Psi_1}\comma \ket{\Psi_2}\big>\equiv \left(\prod_{k=1}^{\ell}\bra{\s_{k; \ell+1-k}}\right)\ket{\Psi_1} \otimes \ket{\Psi_2}\period\label{eq-17}
 } 
Here $\ell$ is the length of the spin chain and $\bra{\s_{k; \ell+1-k}}$ is the state which projects out the singlet part made out of  the spin state at $k$-th site of $\ket{\Psi_1}$ and the one at the  $(\ell+1-k)$-th site of $\ket{\Psi_2}$. The operation should be quite clear from \figref{2pt_comp}. 
In terms of the bracket defined in \eqref{eq-17}, the contraction between $\mathcal{O}_1$ and $\mathcal{O}_2$ can be expressed as
 \beq{
\ewick{\mathcal{O}_1}{\mathcal{O}_2}=\big<\ket{\mathcal{O}_1}_L\comma\ket{\mathcal{O}_2}_L \big>\big<\ket{\tilde{\mathcal{O}}_1}_R\comma\ket{\tilde{\mathcal{O}}_2}_R \big>\period\label{eq-18} 
 }
 As is manifest  in \eqref{eq-18}, the Wick contraction between two operators factorizes into the part coming from the SU(2)$_L$ chain and the part coming from the SU(2)$_R$ chain. This factorization property continues to hold for the tree-level three-point functions, since they are computed through the contractions between the composite operators in the manner described above. 
\section{Construction and evaluation of three-point functions\label{sec:threepnt}}
Up to this point, we have developed a new way of performing the Wick contractions between the composite operators in the spin-chain language suitable for 
dealing with a certain general class of operators in the SU(2) sector. 
We now use this technology to assemble the three-point functions and 
show that they will possess  determinant expressions. 
\subsection{Three-point function as factorized spin-chain products\label{subsec:product}} 
To perform the actual calculations, let us first clarify the basic structure of the 
three-point functions, in particular their characteristic feature of the factorization into the left and the right sector. 

As explained in \cite{Tailoring1},  the three-point function can be computed by first mapping the operators to the spin-chain states, then splitting each spin chain into the left and the right sub-chains (the {\it  cutting} procedure) and finally computing the Wick contractions between the right sub-chain of $\mathcal{O}_1$ and the left sub-chain of $\mathcal{O}_2$ etc.,  using a suitably-chosen inner product for the spin chains (the {\it sewing} procedure). 

 In our formulation, the situation might at first sight appear  more involved,  since each operator $\mathcal{O}_i$ is expressed as a tensor product of two spin-chain states,  $\ket{\mathcal{O}_i}_L$ and $\ket{\tilde{\mathcal{O}}_i}_R$, and then we need to split each of them into two sub-chains. 
However, it is actually more transparent since, as already emphasized,  
 the contributions from the SU(2)$_L$- and SU(2)$_R$-chains completely factorize  and hence  the SU(2)$_L$- and SU(2)$_R$-chains can be discussed separately. Thus, below let us first focus only on the SU(2)$_L$-chain.
 
After the cutting, each spin-chain state is expressed as an entangled state of two states defined on the sub-chains in the following manner:
\beq{
\begin{aligned}
&\ket{\mathcal{O}_1}_L=\sum_a\ket{\mathcal{O}_{1_a}}^{l}\otimes \ket{\mathcal{O}_{1_a}}^{r}\comma\\
&\ket{\mathcal{O}_2}_L=\sum_b\ket{\mathcal{O}_{2_b}}^{l}\otimes \ket{\mathcal{O}_{2_b}}^{r}\comma\\
&\ket{\mathcal{O}_3}_L=\sum_c\ket{\mathcal{O}_{3_c}}^{l}\otimes \ket{\mathcal{O}_{3_c}}^{r}\period
\end{aligned}\label{eq-19}
}
Here the superscripts $l$ and $r$ denote the left and the right sub-chain. The length of each sub-chain is determined from the Wick contraction rule 
and is given by
\beq{
\begin{aligned}
&\text{Length of $\ket{\mathcal{O}_{1_a}}^{r}$ and $\ket{\mathcal{O}_{2_b}}^{l}$}: \frac{\ell_1+\ell_2-\ell_3}{2}\equiv \ell_{12}\comma\\
&\text{Length of $\ket{\mathcal{O}_{2_b}}^{r}$ and $\ket{\mathcal{O}_{3_c}}^{l}$}: \frac{\ell_2+\ell_3-\ell_1}{2}\equiv \ell_{23}\comma\\
&\text{Length of $\ket{\mathcal{O}_{3_c}}^{r}$ and $\ket{\mathcal{O}_{1_a}}^{l}$}: \frac{\ell_3+\ell_1-\ell_2}{2}\equiv\ell_{31}\comma 
\end{aligned}\label{eq-20}
}
where $\ell_i$ is the length of the spin chain $\ket{\calO_i}_L$.  

\begin{figure}[t]
  \begin{center}
   \picture{clip,height=3cm}{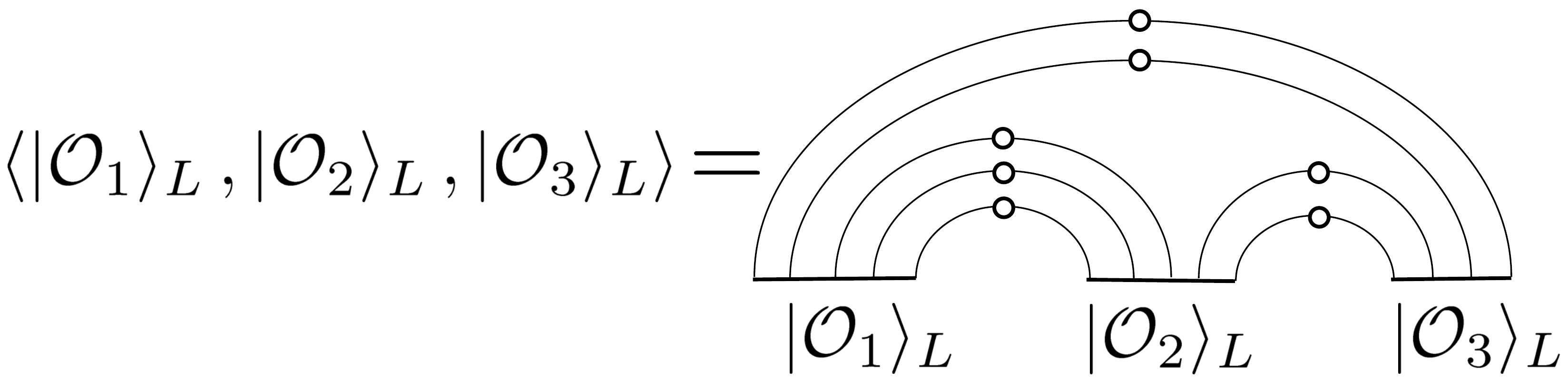}
  \end{center}
\caption{A pictorial definition of the three-spin-chain product. As in \figref{2pt_comp}, at each white blob, we compute the overlap with the singlet state $\bra{\s}$. The number of curves connecting the state $\ket{\mathcal{O}_i}_L$ and $\ket{\mathcal{O}_j}_L$ is determined solely by the length of the operators to be $(\ell_i+\ell_j-\ell_k)/2$.} 
\label{3pt_comp}
\end{figure}

Once the cutting is performed, the rest is to compute the Wick contractions between various sub-chains using the inner product \eqref{eq-17}.  As a result, we get the ``three-spin-chain product'' defined in the following way (see also \figref{3pt_comp}):
\beq{
\big< \ket{\mathcal{O}_1}_L\comma \ket{\mathcal{O}_2}_L \comma\ket{\mathcal{O}_3}_L \big>\equiv \sum_{a,b,c}\big< \ket{\mathcal{O}_{1_a}}^{r}\comma \ket{\mathcal{O}_{2_b}}^{l}\big>\big< \ket{\mathcal{O}_{2_b}}^{r}\comma \ket{\mathcal{O}_{3_c}}^{l}\big>\big< \ket{\mathcal{O}_{3_c}}^{r}\period \ket{\mathcal{O}_{1_a}}^{l}\big>\period\label{eq-21}
}
Multiplying the contribution from the SU(2)$_R$ sector, which is 
entirely similar to \eqref{eq-21},  the final formal expression for the structure constant is given by
\beq{
C_{123}=\frac{\sqrt{\ell_1 \ell_2 \ell_3}}{N_c\sqrt{\mathcal{N}_1\mathcal{N}_2\mathcal{N}_3}}\big< \ket{\mathcal{O}_1}_L\comma \ket{\mathcal{O}_2}_L \comma\ket{\mathcal{O}_3}_L \big>\big< \ket{\tilde{\mathcal{O}}_1}_R\comma \ket{\tilde{\mathcal{O}}_2}_R \comma\ket{\tilde{\mathcal{O}}_3}_R \big>\comma \label{eq-22}
}
where $\mathcal{N}_k$ denotes a factor coming from the normalization of the operator $\mathcal{O}_k$.
As advertised several times already, the expression \eqref{eq-22} 
of the structure constant completely factorizes into the contributions 
 from the SU(2)$_L$ and the SU(2)$_R$ parts. 
This phenomenon was already observed in \cite{Tailoring1,Foda} for a restricted class of three-point functions but \eqref{eq-22} tells us that it is a much more general  property as long as three SU(2)-operators can be embedded in a single SO(4). 
In any case, the expression of the structure constant above is as yet formal, and in the rest of this  section we shall perform the cutting and sewing explicitly 
in our new formalism and that will naturally lead to the determinantal formula  for the three-point functions. 
\subsection{``Cutting and sewing" in the new formulation\label{subsec:cutandsew}}
Let us begin with the explanation of the cutting procedure in our formalism. 
Due to the factorization property we only need to focus on the SU(2)$_L$ part. 
Below we only consider the operators satisfying the highest weight conditions. As shown in \eqref{eq-13}, such 
operators can be expressed as Bethe states multiplied by the operator $e^{zS_{-}}$. The cutting procedure of Bethe states is already studied in \cite{Tailoring1} using the method called ``generalized two-component model'' and the result in our notation takes the form 
\beq{
\ket{\bm{u};\upspin^{\ell}}_L=\sum_{\bm{\alpha}_l\cup\bm{\alpha}_r=\bm{u}}H_{\ell}(\bm{\alpha}_l\comma \bm{\alpha}_r|\bm{\theta})\ket{\bm{\alpha}_{l};\upspin^{\ell_l}}\otimes\ket{\bm{\alpha}_{r};\upspin^{\ell_r}}\period\label{eq-35}
}
The sum is over all possible ways of splitting the rapidities $\bm{u}$ into two groups $\bm{\alpha}_l$ and $\bm{\alpha}_r$, the symbols $\ell_l$ and $\ell_r$ denote respectively the length of the left and the right sub-chains\fn{Note that $\ell_l$ and $\ell_r$ satisfy $\ell= \ell_l+\ell_r$.} and the coefficient function $H_{\ell}(\bm{\alpha}_l\comma \bm{\alpha}_r|\bm{\theta})$ reads\footnote{Just as in \cite{Tailoring1}, this coefficient is obtained by re-expressing  $B(u)$ of the original chain in terms  the elements of the monodromy matrices $\Omega_l$ and $\Omega_r$ of the left and the right sub-chains through the relation $B(u)=\Omega(u)_{12} = (\Omega_l(u) \Omega_r(u))_{12}$ and then pushing the operators $A_{l}$ and $D_r$ to the right using the Yang-Baxter algebra.}
\beq{
&H_{\ell}(\bm{\alpha}_l\comma \bm{\alpha}_r|\bm{\theta})\equiv \prod_{u\in \bm{\alpha}_l}\prod_{v\in\bm{\alpha}_r}\prod_{a=\ell_l+1}^{\ell}\prod_{b=1}^{\ell_l}\left(\frac{u-v+i}{u-v}\right)\left(u-\theta_a -\frac{i}{2}\right)\left(v-\theta_b +\frac{i}{2}\right)\period\label{eq-36}
}
On the other hand, the splitting of the prefactor $e^{zS_{-}}$ is simple since the global SU(2) generator $S_{-}$ for the full chain is just a sum of the generators for the sub-chains: $S_{-}=S_{-}^{l}\otimes \bm{1} +\bm{1}\otimes S_{-}^{r}$. Therefore, after the cutting procedure, the rotated excited state \eqref{eq-13} is expressed as
\beq{
e^{z S_{-}}\ket{\bm{u};\upspin^{\ell}}_L=\sum_{\bm{\alpha}_l\cup\bm{\alpha}_r=\bm{u}}H_{\ell}(\bm{\alpha}_l\comma \bm{\alpha}_r)\Big(e^{zS_{-}^{l}} \ket{\bm{\alpha}_{l};\upspin^{\ell_l}}\Big)\otimes\Big( e^{zS_{-}^{r}}\ket{\bm{\alpha}_{r};\upspin^{\ell_r}}\Big)\period\label{modifiedsplit}
}

Although the cutting procedure described above is quite similar to the one 
 developed in \cite{Tailoring1}, except for the SU(2)$_L$-SU(2)$_R$  factorization 
property, the sewing procedure in our formalism is substantially different, 
 with a definite advantage. To describe this, we use an important 
property, which we call the ``crossing'' relation, of the Lax operator
\beq{
{\rm L}(u)=\pmatrix{cc}{u+iS_3&iS_{-}\\iS_{+}&u-iS_3}\period\label{Laxdef}
}
Let $\ket{s_1}$ and $\ket{s_2}$ be two arbitrary spin 1/2 states and consider the overlap with the singlet state with the Lax-operator insertion: $\bra{\s}\left( {\rm L}(u-\theta)\ket{s_1}\otimes \ket{s_2}\right)$. Using the definition of the singlet state, one can show the following relation by direct computation:
\beq{
\bra{\s}\left( {\rm L}(u-\theta)\ket{s_1}\otimes \ket{s_2}\right)=\bra{\s}\left( \ket{s_1}\otimes \mathcal{C}\circ{\rm L}(u-\theta)\ket{s_2}\right)\comma\label{crossing1}
 } 
 where $\mathcal{C}\circ {\rm L}(u)$ is the ``crossed Lax operator'', which is given by
 \beq{
 \mathcal{C}\circ {\rm L}(u) \equiv \sigma_2 {\rm L}^{t}(u)\sigma_2 =\pmatrix{cc}{u-iS_{3}&-iS_{-}\\-iS_{+}&u+iS_{3}}\period\label{crossed1}
 }
In \eqref{crossed1}, $\sigma_2$ acts on the auxiliary space and the superscript $t$ denotes the transposition in the auxiliary space. This relation, if we regard $\sigma_2$ as the charge conjugation matrix, can be viewed as a sort of the crossing relation of the factorized S-matrices\fn{For the relation between the crossing symmetry and the scattering with the singlet state, see, for example, \cite{Beisert,dressing}. See also the footnote \ref{fn:crossing}.} and we therefore call \eqref{crossed1} the ``crossing'' relation.  

\begin{figure}[t]
\begin{minipage}{0.5\hsize}
    \begin{center}
   \picture{clip,height=3cm}{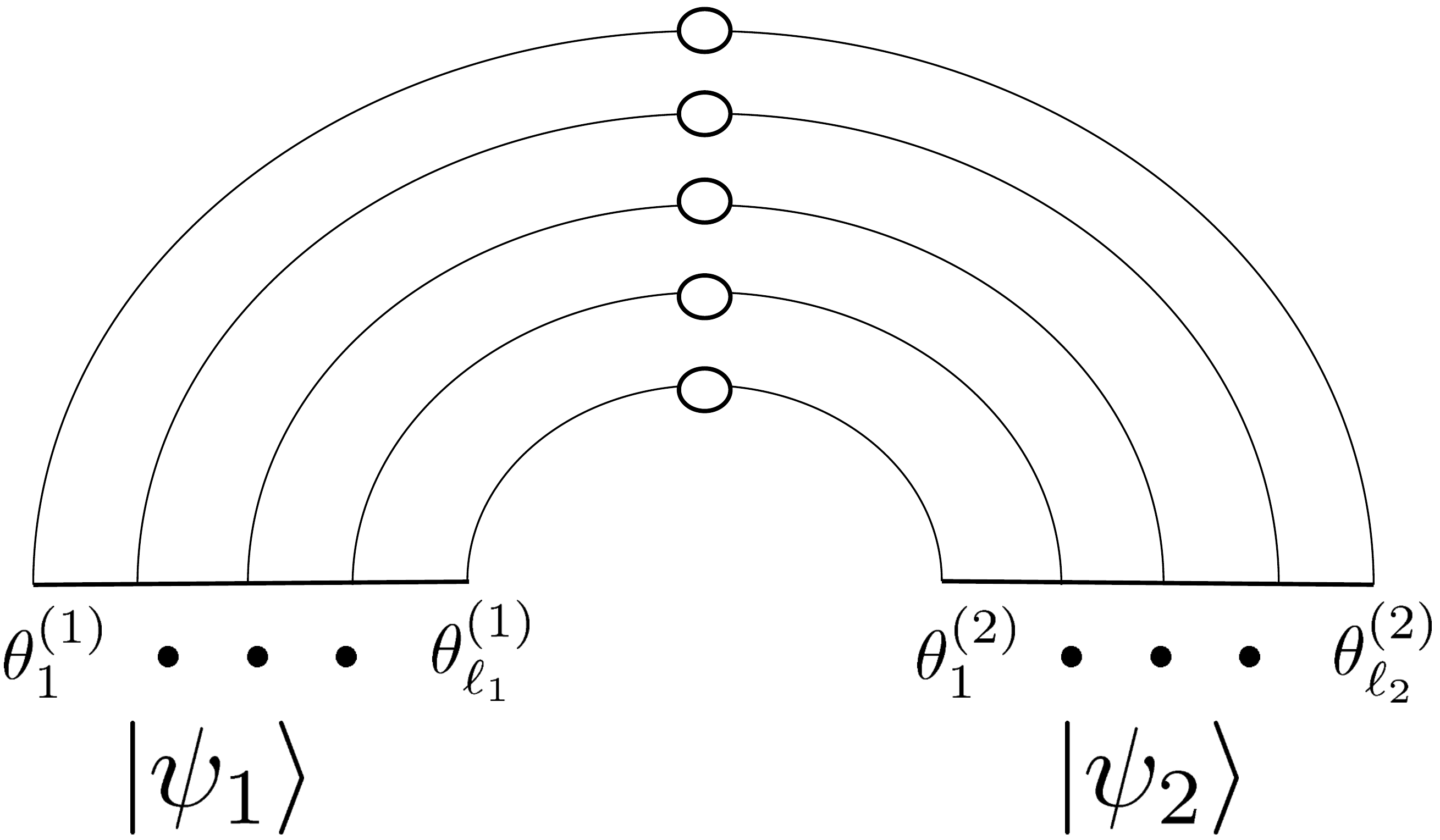}\\
   (a)
  \end{center}
  \end{minipage}
\begin{minipage}{0.5\hsize}
    \begin{center}
   \picture{clip,height=3cm}{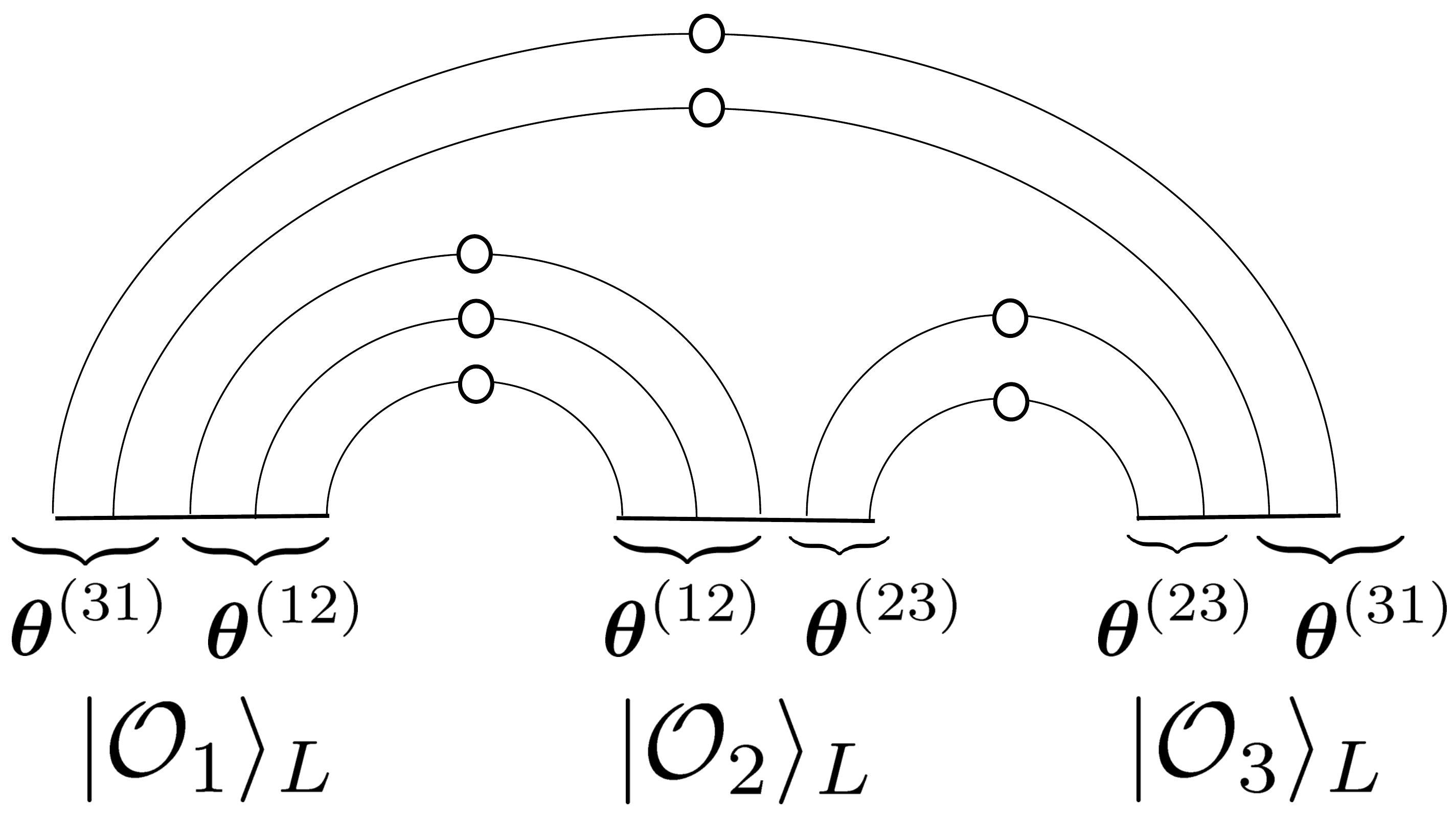}\\
   (b)
  \end{center}
  \end{minipage}
\caption{The identification of the inhomogeneities for the two-point functions and the three-point functions. In both cases, we identify the inhomogeneities connected by a propagator (connected to the same blob). (a) For the two-point functions, the identification is given by \eqref{inhomoeq}. (b) For the three-point functions, the sets of the inhomogeneities are related as \eqref{setinhomo}.} 
\label{inhomo}
\end{figure}

The relation \eqref{crossed1} leads to a useful nontrivial identity of the monodromy matrix. Let $\ket{\psi_1}$ and $\ket{\psi_2}$ to be arbitrary spin-chain states of the same length. They can be either on-shell Bethe states describing each operator or off-shell Bethe states which appear after the cutting procedure. Then, from the fundamental relation \eqref{crossed1},  the following important relation can be obtained, as we shall prove shortly:
\beq{
\big< \Omega_1 (u) \ket{\psi_1}\comma \ket{\psi_2}\big>=\big<  \ket{\psi_1}\comma \sigma_2\, \Omega_2^{t} (u)\,\sigma_2 \ket{\psi_2}\big>\period\label{eq-37}
 }
 Here, again  $t$ and $\sigma_2$ act on the auxiliary space and  $\Omega_{n}$ is the monodromy matrix acting on $\ket{\psi_n}$ defined by
 \beq{
 \begin{aligned}
 &\Omega_1(u) = {\rm L}_1(u-\theta^{(1)}_1)\cdots {\rm L}_{\ell}(u-\theta^{(1)}_\ell)\comma\quad \Omega_2(u) = {\rm L}_1(u-\theta^{(2)}_1)\cdots {\rm L}_{\ell}(u-\theta^{(2)}_\ell)\period
  \end{aligned}\label{defombra}
  } 
 The parameters $\theta^{(n)}$'s are the inhomogeneities for $\ket{\psi_n}$. In order for \eqref{eq-37} to be satisfied, we need to make the following identification between the inhomogeneities (see \figref{inhomo}):
  \beq{
  \theta^{(1)}_k = \theta^{(2)}_{\ell-k+1}\period\label{inhomoeq}
  }
In terms of the Wick contraction in the gauge theory, this amounts to assigning the same inhomogeneity parameter to each two spin sites contracted by a propagator. This is precisely the identification we need when we study the one-loop correction using the inhomogeneities \cite{Tailoring4, Kostov1, Kostov2, Fixing} and we impose such relation throughout this paper.

Let us now prove the relation \eqref{eq-37}. From the definition of $\Omega$ \eqref{defombra}, we can express the LHS of \eqref{eq-37} as
 \beq{
 \big< \left( {\rm L}_1(u-\theta^{(1)}_1)\right)_{i_1i_2}\left({\rm L}_2(u-\theta^{(1)}_2)\right)_{i_2i_3}\cdots \left( {\rm L}_{\ell}(u-\theta^{(1)}_{\ell})\right)_{i_\ell i_{\ell+1}} \ket{\psi_1}\comma \ket{\psi_2}\big>\comma
 }
 where, for definiteness, we wrote down the indices for the auxiliary space.
 Since the $k$-th site of $\ket{\psi_1}$ is contracted with the $(\ell-k+1)$-th site of $\ket{\psi_2}$ as shown in \figref{inhomo} and the inhomogeneities are identified as \eqref{inhomoeq}, the Lax operator transforms as follows under the application of the crossing relation \eqref{crossed1}:
 \beq{
 \left( {\rm L}_k (u-\theta^{(1)}_k)\right)_{i_ki_{k+1}} \to \quad \left( \sigma_2 {\rm L}^{t}_{\ell-k+1}(u-\theta^{(2)}_{\ell-k+1})\sigma_2\right)_{i_k i_{k+1}}\period
 }
 Then, moving the Lax operators one by one, we obtain
 \beq{ 
 \begin{aligned}
&\big<  \ket{\psi_1}\comma \left( \sig_2 {\rm L}_{\ell}^t(u-\theta^{(2)}_{\ell}) \sig_2\right)_{i_1i_2} 
 \left( \sig_2 {\rm L}_{\ell-1}^t(u-\theta^{(2)}_{\ell-1}) \sig_2\right)_{i_2i_3}\cdots \left( \sig_2 {\rm L}_{1}^t(u-\theta^{(2)}_{1}) \sig_2\right)_{i_{\ell} i_{\ell+1}} \ket{\psi_2}\big> \\
& = \big< \ket{\psi_1} \comma \left( \sig_2 ( {\rm L}_{1}(u-\theta^{(2)}_{1})\cdots {\rm L}_{\ell}(u-\theta^{(2)}_{\ell}) )^t \sig_2\right)_{i_1 i_{\ell+1}} \ket{\psi_2} \big>\\
&= \big<  \ket{\psi_1}\comma \left( \sigma_2  \Omega_2 ^t(u)\sigma_2\right)_{i_1i_{\ell+1}} \ket{\psi_2}\big>\period
\end{aligned}
\label{eq-38}
 }
In terms of components, $\sigma_2  \Omega^t_2(u) \sigma_2$ is given by
  \beq{
 \sigma_2\Omega_{2}^t (u)\sigma_2=\sigma_2\pmatrix{cc}{A^{(2)}(u)&B^{(2)}(u)\\C^{(2)}(u)&D^{(2)}(u)}^t\sigma_2=\pmatrix{cc}{D^{(2)}(u)&-B^{(2)}(u)\\-C^{(2)}(u)&A^{(2)}(u)}\period
  }
Here and throughout this subsection we put superscripts $(1)$ or $(2)$ in order to distinguish the components of $\Omega_1$ from those of $\Omega_2$.
The formula \eqref{eq-37} in particular contains the crucial relation
  \beq{
  \big<B^{(1)}(u)\ket{\psi_1}\comma \ket{\psi_2}\big>=-\big<\ket{\psi_1}\comma B^{(2)}(u)\ket{\psi_2}\big>\comma\label{eq-39}
  }
which only involves $B(u)$ operators. In the $u\to\infty$ limit, the relation \eqref{eq-39} produces
\beq{
\big<S^{(1)}_{-}\ket{\psi_1}\comma \ket{\psi_2}\big>=-\big<\ket{\psi_1}\comma S^{(2)}_{-}\ket{\psi_2}\big>\period\label{s-relation}
 } 
 Then, by a repeated use of \eqref{eq-39} and \eqref{s-relation}, we can collect all $B(u)$'s and $S_{-}$'s on one side and transform the skew-symmetric inner product which appear in the sewing procedure, such as 
 \beq{
 \big< e^{xS^{(1)}_{-}}B^{(1)}(u_1)\cdots B^{(1)}(u_{M_1})\ket{\upspin^{\ell}}\comma e^{yS^{(2)}_{-}}B^{(2)}(v_1)\cdots B^{(2)}(v_{M_2})\ket{\upspin^{\ell}}\big>  \comma 
 }
into the following expression: 
 \beq{
(-1)^{M_1}\big<  \ket{\upspin^{\ell}}\comma e^{(y-x)S^{(2)}_{-}}B^{(2)}(u_1)\cdots B^{(2)}(u_{M_1})B^{(2)}(v_1)\cdots B^{(2)}(v_{M_2})\ket{\upspin^{\ell}}\big>\period\label{eq-40}
 }

From the definition of the skew-symmetric inner product \eqref{eq-17}, this expression can be readily evaluated\footnote{Essentially, due to the 
skew-symmetry,  each time  the singlet projector acts on a pair of spins, the up-spin is converted to the down-spin and this produces $\bra{\downarrow^\ell\!\!}$.}
 as a matrix element in the spin-chain Hilbert space as follows: 
 \beq{
 \begin{aligned}
 &(-1)^{M_1}\big<  \ket{\upspin^{\ell}}\comma e^{(y-x)S^{(2)}_{-}} B^{(2)}(u_1)\cdots B^{(2)}(u_{M_1})B^{(2)}(v_1)\cdots B^{(2)}(v_{M_2})\ket{\upspin^{\ell}}\big>\\
 & =(-1)^{M_1}\bra{\downarrow^{\ell}\!}e^{(y-x)S^{(2)}_{-}}B^{(2)}(u_1)\cdots B^{(2)}(u_{M_1})B^{(2)}(v_1)\cdots B^{(2)}(v_{M_2})\ket{\upspin^{\ell}}\period 
 \end{aligned}\label{eq-41}
 }
It is important to recognize that a matrix element of the form \eqref{eq-41} can be identified with the so-called partial domain wall partition function. More precisely, we can show
 \beq{
\bra{\downarrow^{\ell}\!}e^{zS_{-}}B(x_1)\cdots B(x_{M})\ket{\upspin^{\ell}}=z^{\ell-M}Z_{p}(\bm{x}|\bm{\theta})\comma\label{funda}
}
where $Z_{p}(\bm{x}|\bm{\theta})$ is the partial domain wall partition function (pDWPF), which is given by  \cite{Kostov1,Kostov2}:
\beq{
 \begin{aligned}
 Z_{p}\left(\bm{x}|\bm{\theta}\right)& \equiv{\frac{1}{(\ell-M)!}}\bra{\downarrow^{\ell}\!}S_{-}^{\ell-M}B(x_1)\cdots B(x_{M})\ket{\upspin^{\ell}}\\
 &=\frac{\prod_{i=1}^{M}\prod_{j=1}^{\ell} (x_i-\theta_j-i/2)}{\prod_{i<j}(x_i-x_j)}\det \left(x_b^{a-1}-\prod_{c=1}^{\ell}\frac{x_b-\theta_c+i/2}{x_b-\theta_c-i/2}(x_b-i)^{a-1}\right)_{ a,b}\period
 \end{aligned}
 \label{eq-43}
 }
The indices $a$ and $b$ run from $1$ to $M$ and $\theta$'s are the inhomogeneity parameters for the chain.
To understand \eqref{funda}, we just need to expand the exponential $e^{z S_{-}}$ on the LHS of \eqref{funda}. Upon doing so, \eqref{funda} yields infinitely many terms, each of which has a different number of $S_{-}$'s. However, among such terms, only one term  
\beq{
\frac{z^{\ell-M}}{(\ell-M)!}\bra{\downarrow^{\ell}\!}\left( S_{-}\right)^{\ell-M} B(x_1)\cdots B(x_{M})\ket{\upspin^{\ell}}\label{oneterm}
}
is non-vanishing because of the conservation of the SU(2) spin and it can be readily identified with the pDWPF.

 Let us stress that the discussion above is valid both for the on-shell and the off-shell Bethe states. In \cite{KM}, it was shown that the scalar product between the on-shell Bethe state and the off-shell Bethe state can be transformed into the pDWPF\fn{In fact, using our formulation, one can prove the equivalence between the pDWPF and the on-shell-off-shell scalar product (the so-called Kostov-Matsuo trick) by a simple calculation.}. However, such an argument cannot be applied to the scalar products between two off-shell Bethe states and this was considered to be the main obstacle in studying more general SU(2) three-point functions. In this respect, the argument above clearly shows the advantage of our formulation based on the skew-symmetric inner product as it allows us to use the determinant expression irrespective of whether the Bethe states are on-shell or not. 
\subsection{Representation in terms of the partial domain wall partition function\label{subsec:pDWPF}}
Let us now combine the results in the previous subsections to write down an explicit expression for general three-point functions. As in the previous subsections, we focus on the contribution from the SU(2)$_L$ sector $\big< \ket{\mathcal{O}_1}_L\comma \ket{\mathcal{O}_2}_L\comma \ket{\mathcal{O}_3}_L\big>$. 

First, using the coset parametrization \eqref{eq-13}, each spin-chain state can be expressed as
\beq{
\begin{aligned}
&\ket{\mathcal{O}_1}_L=\left( \frac{1}{1+|z_1|^2}\right)^{\ell_1/2 -M_1}e^{z_1 S_{-}}\ket{\bm{u}^{(1)};\upspin^{\ell_1}}\comma\\
&\ket{\mathcal{O}_2}_L=\left( \frac{1}{1+|z_2|^2}\right)^{\ell_2/2 -M_2}e^{z_2 S_{-}}\ket{\bm{u}^{(2)};\upspin^{\ell_2}}\comma\\
&\ket{\mathcal{O}_3}_L=\left( \frac{1}{1+|z_3|^2}\right)^{\ell_3/2 -M_3}e^{z_3 S_{-}}\ket{\bm{u}^{(3)};\upspin^{\ell_3}}\comma
\end{aligned}\label{3ptstates}
}
where  $\bm{u}^{(k)}$ denotes the set of  rapidities for the operator $\mathcal{O}_k$ and its number of elements is denoted by $M_k$. Then, we can apply the formula \eqref{eq-35} to split each chain into two and compute the skew-symmetric inner product using \eqref{funda}. When computing the inner product, it is important that we assign the same inhomogeneity parameter to any two spin sites contracted by a propagator as discussed in the previous subsection. In the current setup, this leads to the following relation among the sets of inhomogeneities (see \figref{inhomo}):
\beq{
\bm{\theta}^{(1)}=\bm{\theta}^{(31)}\cup\bm{\theta}^{(12)}\comma \quad \bm{\theta}^{(2)}=\bm{\theta}^{(12)}\cup\bm{\theta}^{(23)}\comma \quad \bm{\theta}^{(3)}=\bm{\theta}^{(23)}\cup\bm{\theta}^{(31)}\comma\label{setinhomo}
  }  
  where $\bm{\theta}^{(n)}$ is the set of inhomogeneities for $\ket{\mathcal{O}_n}_L$ and $\bm{\theta}^{(nm)}$ denote the set of the inhomogeneities common to $\ket{\mathcal{O}_n}_L$ and $\ket{\mathcal{O}_m}_L$.
As a result of these operations, we obtain the following final form 
expressed in terms of  the sum-over-partitions
\beq{
\begin{aligned}
&\big< \ket{\mathcal{O}_1}_L\comma \ket{\mathcal{O}_2}_L\comma \ket{\mathcal{O}_3}_L\big>\\
&=\left( \frac{1}{1+|z_1|^2}\right)^{\ell_1/2 -M_1}\left( \frac{1}{1+|z_2|^2}\right)^{\ell_2/2 -M_2}\left( \frac{1}{1+|z_3|^2}\right)^{\ell_3/2 -M_3}\\
&\qquad\times \sum_{\bm{\alpha}_l^{(k)}\cup\bm{\alpha}_r^{(k)}=\bm{u}^{(k)} } z_{21}^{\ell_{12}-|\bm{\alpha}_r^{(1)}|-|\bm{\alpha}_l^{(2)}|}z_{32}^{\ell_{23}-|\bm{\alpha}_r^{(2)}|-|\bm{\alpha}_l^{(3)}|}z_{13}^{\ell_{31}-|\bm{\alpha}_r^{(3)}|-|\bm{\alpha}_l^{(1)}|} \, \mathcal{D}_{\{\bm{\alpha}_{l,r}^{(1)},\bm{\alpha}_{l,r}^{(2)}\bm{\alpha}_{l,r}^{(3)}\}}\comma\label{generalformula}
 \end{aligned}
 } 
 In this expression,  $|\bm{\alpha}_{l,r}^{(k)}|$  stands for the number of elements of $\bm{\alpha}_{l,r}^{(k)}$ and $z_{nm}$ denotes the difference $z_n-z_m$. The last factor $\mathcal{D}_{\{\bm{\alpha}_{l,r}^{(1)},\bm{\alpha}_{l,r}^{(2)}\bm{\alpha}_{l,r}^{(3)}\}}$, which is independent
of the polarizations, is given in terms of the pDWPF as
 \beq{
 \begin{aligned}
 \mathcal{D}_{\{\bm{\alpha}_{l,r}^{(1)},\bm{\alpha}_{l,r}^{(2)}\bm{\alpha}_{l,r}^{(3)}\}}&\equiv (-1)^{|\bm{\alpha}_r^{(1)}|+|\bm{\alpha}_r^{(2)}|+|\bm{\alpha}_r^{(3)}|}\prod_{k=1}^{3}H_{\ell_k}(\bm{\alpha}_{l}^{(k)},\bm{\alpha}_{r}^{(k)}|\bm{\theta}^{(k)})\\
 &\times Z_p\left(\bm{\alpha}_r^{(1)}\cup\bm{\alpha}_l^{(2)}|\bm{\theta}^{(12)} \right)Z_p\left(\bm{\alpha}_r^{(2)}\cup\bm{\alpha}_l^{(3)}|\bm{\theta}^{(23)}\right)Z_p\left(\bm{\alpha}_r^{(3)}\cup\bm{\alpha}_l^{(1)}|\bm{\theta}^{(31)}\right)\period 
\end{aligned}
  }
 
Let us emphasize that our final expression \eqref{generalformula} has a number of advantages.  Firstly, the result is valid for the three-point functions built upon more general spin-chain vacua than the ones studied in the literature. Secondly, the result already demonstrates certain separation into the kinematical factor and the dynamical factor. Thirdly , the dynamical factor $\mathcal{D}_{\{\bm{\alpha}_{l,r}^{(1)},\bm{\alpha}_{l,r}^{(2)}\bm{\alpha}_{l,r}^{(3)}\}}$ is given essentially by a product of the pDWPF, each of which possesses  determinant representation.  
One apparently unsatisfactory  feature of \eqref{generalformula}  is that it still involves the sums over partitions, which become quite nontrivial especially when the number of  magnons is  large. As we shall show in the next subsection,  however,  for certain class of correlators  the sum can be 
 reduced to  just a single term,  by exploiting the SU(2) symmetry\fn{A similar idea was utilized to simplify the three-point functions in the SL(2) sector in \cite{non-compact}.}. This 
 leads to a remarkably simple expression for which the semi-classical limit can be easily taken.
\subsection{Determinant expressions for a large  class of three-point functions\label{subsec:det}}
 Let us now show that for certain correlators the expression \eqref{generalformula} can be drastically  simplified. The correlators we consider are those for which two of the operators belong to one type (type I or type II) and the third
 to the other type. We call such three-point functions ``mixed correlators''. In what follows, we  study the case in which $\mathcal{O}_1$ and $\mathcal{O}_2$ are of type I and $\mathcal{O}_3$ is of type II since the generalization to other cases is simply a matter of renaming. 

The crucial observation for the simplification is the fact that  the  dependence on the parameters $z_i$ characterizing the operators forming  the three-point functions is completely dictated by the SU(2) symmetry. This is shown 
 in Appendix \ref{ap} and is quite analogous to the determination of the 
position dependence for the three-point functions in two-dimensional conformal field theory.  Now as $\mathcal{O}_3$ is of type II and therefore $\ket{\mathcal{O}_3}_L$ contains  no magnons in the present case, we can set $M_3$ in the formula \eqref{kineL} to zero and obtain
 \beq{
 \begin{aligned}
 \big<\ket{\mathcal{O}_1}_L\comma\ket{\mathcal{O}_2}_L\comma \ket{\mathcal{O}_3}_L \big>=& \left( \frac{1}{1+|z_1|^2}\right)^{\frac{\ell_1}{2} -M_1}\left( \frac{1}{1+|z_2|^2}\right)^{\frac{\ell_2}{2} -M_2}\left( \frac{1}{1+|z_3|^2}\right)^{\frac{\ell_3}{2}}\\
&\times z_{21}^{\ell_{12}-M_1-M_2}z_{32}^{\ell_{23}-M_2+M_1}z_{13}^{\ell_{31}-M_1+M_2}\mathcal{G}\comma\label{structuremixed}
  \end{aligned}
    }
    where the factor $\mathcal{G}$ stands for the  term independent of $z_i$'s.  As can be  easily seen, the first line of \eqref{structuremixed} coincides with the second line of \eqref{generalformula}. On the other hand, the structure given in the second line of \eqref{structuremixed} is not visible in the sum-over-partition expression \eqref{generalformula}. In order to compare them more closely, let us expand both sides in powers of $z_3$. Upon this expansion, the second line of \eqref{structuremixed} yields the following term as the highest-order term:
    \beq{
    (-1)^{\ell_{31}-M_1+M_2}z_3^{\ell_3}\left( z_{21}^{\ell_{12}-M_1-M_2}\mathcal{G}\right)\period
    }
    On the other hand, if we expand each term in the sum in \eqref{generalformula}, we obtain the following expression as the highest-order term\fn{Note that, since $\ket{\mathcal{O}_3}_L$ does not have any magnons, there is no sum over the partitions coming from $\mathcal{O}_3$.}:
    \beq{
    (-1)^{\ell_{31}-|\bm{\alpha}_l^{(1)}|}z_{3}^{\ell_3-|\bm{\alpha}_l^{(1)}|-|\bm{\alpha}_r^{(2)}|}\left( z_{21}^{\ell_{12}-|\bm{\alpha}_r^{(1)}|-|\bm{\alpha}_l^{(2)}|}\right)\mathcal{D}_{\{\bm{\alpha}_{l,r}^{(1)},\bm{\alpha}_{l,r}^{(2)}, \varnothing\}}\period
    }
    This shows that only a single term in the sum, for which  $|\bm{\alpha}_l^{(1)}|=|\bm{\alpha}_r^{(2)}|=0$ holds, can produce the highest power $z_3^{\ell_3}$. Therefore, comparing the coefficients in front of $z_3^{\ell_3}$, we can determine $\mathcal{G}$ to be of the form 
\beq{
\begin{aligned}
\mathcal{G}&=\left.(-1)^{-M_1+M_2}\mathcal{D}_{\{\bm{\alpha}_{l,r}^{(1)},\bm{\alpha}_{l,r}^{(2)}, \varnothing\}}\right|_{\bm{\alpha}_l^{(1)}=\bm{\alpha}_r^{(2)}=\varnothing}\\
&=(-1)^{M_2}\prod_{a=1}^{M_1}Q_{\bm{\theta}^{(31)}}^{+}(u_a^{(1)})\prod_{b=1}^{M_2}Q_{\bm{\theta}^{(23)}}^{-}(u_b^{(2)})Z_p\left( \bm{u}^{(1)}\cup \bm{u}^{(2)}|\bm{\theta}^{(12)}\right)\comma 
\end{aligned}
}
where  the function $Q_{\bm{\theta}}(x)$ is defined by 
\beq{
Q_{\bm{\theta}}(x) \equiv \prod_{\theta \in \bm{\theta}}(x-\theta)\comma
}
and the superscripts $\pm$ denote the shift of the argument by $\pm i/2$. 

Let us now study the semi-classical limit of our three-point function.  For this 
purpose,  it is more convenient to introduce the ``rescaled" partial domain wall partition function\fn{Note that it is the rescaled partial domain wall partition function, which has a simple semi-classical limit. In \cite{Tailoring3}, it is called $\mathcal{A}$-functional.} defined by
\beq{
\mathcal{Z}_p\left(\bm{u}^{(1)}\cup\bm{u}^{(2)}|\bm{\theta}^{(12)}\right) \equiv \frac{Z_p\left(\bm{u}^{(1)}\cup\bm{u}^{(2)}|\bm{\theta}^{(12)}\right)}{\prod_{x\in \bm{u}^{(1)}}Q_{\bm{\theta}^{(12)}}^{+}(x)\prod_{y\in \bm{u}^{(2)}}Q_{\bm{\theta}^{(12)}}^{-}(y)}\period
 }
Then, $\big<\ket{\mathcal{O}_1}_L\comma\ket{\mathcal{O}_2}_L\comma \ket{\mathcal{O}_3}_L \big>$ takes the form 
\beq{
\begin{aligned}
\big<\ket{\mathcal{O}_1}_L\comma \ket{\mathcal{O}_2}_L\comma \ket{\mathcal{O}_3}_L\big>=&\left( \frac{1}{1+|z_1|^2}\right)^{\frac{\ell_1}{2} -M_1}\left( \frac{1}{1+|z_2|^2}\right)^{\frac{\ell_2}{2} -M_2}\left( \frac{1}{1+|z_3|^2}\right)^{\frac{\ell_3}{2} }\\
&\times(z_1-z_2)^{\ell_{12}-M_1-M_2}(z_2-z_3)^{\ell_{23}-M_2+M_1}(z_3-z_1)^{\ell_{31}-M_{1}+M_2}\\
&\times\left(\prod_{a=1}^{M_1}Q^{+}_{\bm{\theta}^{(1)}}(u_a^{(1)})\prod_{b=1}^{M_2}Q^{-}_{\bm{\theta}^{(2)}}(u_b^{(2)})\mathcal{Z}_p\left(\bm{u}^{(1)}\cup\bm{u}^{(2)}|\bm{\theta}^{(12)}\right)\right)\comma
\end{aligned}\label{resultleft}
}
where we have neglected the factor $(-1)^{M_2}$ as it only changes the overall sign.
Performing a similar analysis, we can also determine the contribution from the SU(2)$_R$ spin chain and the result is given by 
\beq{
\begin{aligned}
\big<\ket{\tilde{\mathcal{O}}_1}_R\comma \ket{\tilde{\mathcal{O}}_2}_R\comma \ket{\tilde{\mathcal{O}}_3}_R\big>=&\left( \frac{1}{1+|\tilde{z}_1|^2}\right)^{\frac{\ell_1}{2} }\left( \frac{1}{1+|\tilde{z}_2|^2}\right)^{\frac{\ell_2}{2}}\left( \frac{1}{1+|\tilde{z}_3|^2}\right)^{\frac{\ell_3}{2} -\tilde{M}_3}\\
&\times(\tilde{z}_1-\tilde{z}_2)^{\ell_{12}+\tilde{M}_3}(\tilde{z}_2-\tilde{z}_3)^{\ell_{23}-\tilde{M}_3}(\tilde{z}_3-\tilde{z}_1)^{\ell_{31}-\tilde{M}_{3}}\\
&\times\left(\prod_{a=1}^{M_3}Q^{+}_{\tilde{\bm{\theta}}^{(3)}}(\tilde{u}_a^{(3)})\mathcal{Z}_p\left( \tilde{\bm{u}}^{(3)}\cup\varnothing|\tilde{\bm{\theta}}^{(31)}\right)\right)\period
\end{aligned}\label{resultright}
}
Note that, since the result  is completely factorized into the SU(2)$_L$ and the SU(2)$_R$ parts, we can introduce independent  sets of the inhomogeneities for the SU(2)$_R$ sector  denoted by $\tilde{\theta}$'s.
The tree-level structure constant can then be obtained by setting $\theta$'s and $\tilde{\theta}$'s to zero. Now the  semi-classical limit of the three-point 
 coupling constant can also be easily studied  using the results of \cite{Tailoring3,Kostov1,Kostov2} and we obtain, 
 up to a phase, 
\beq{
\begin{aligned}
C_{123}&=\frac{\sqrt{\ell_1\ell_2\ell_3}}{N_c}k_L \, k_R\,  c_{123}\comma\\
\log c_{123} &\sim  \oint_{\mathcal{C}_{\bm{u}^{(1)}}\cup \mathcal{C}_{\bm{u}^{(2)}}}\frac{du}{2\pi i} {\rm Li}_2 \left(e^{ip_1+ip_2 +i\ell_3/2u} \right)+\oint_{\mathcal{C}_{\tilde{\bm{u}}^{(3)}}}\frac{du}{2\pi i} {\rm Li}_2 \left(e^{ip_3 +i(\ell_2-\ell_1)/2u} \right)\\
&-\frac{1}{2}\oint_{\mathcal{C}_{\bm{u}^{(1)}}}\frac{du}{2\pi} {\rm Li}_2 \left( e^{2i p_1}\right)-\frac{1}{2}\oint_{\mathcal{C}_{\bm{u}^{(2)}}}\frac{du}{2\pi} {\rm Li}_2 \left( e^{2i p_2}\right)-\frac{1}{2}\oint_{\mathcal{C}_{\tilde{\bm{u}}^{(3)}}}\frac{du}{2\pi} {\rm Li}_2 \left( e^{2i \tilde{p}_3}\right)\period
\end{aligned}
}
Here  $k_L$ and $k_R$ are kinematical factors given by the first two lines on the left hand side of \eqref{resultleft} and \eqref{resultright} respectively, $p_{n}(u)$ and $\tilde{p}_{n} (u)$ are the quasi-momenta given by
\beq{
p_{n}(u) = \sum_{v\in \bm{u}^{(n)}} \frac{1}{u-v}- \frac{\ell_n}{2u}\comma\qquad \tilde{p}_{n}(u) = \sum_{v\in \tilde{\bm{u}}^{(n)}} \frac{1}{u-v}- \frac{\ell_n}{2u}\comma 
}
and the integration contours $\mathcal{C}_{\bm{u}^{(n)}}$ and $\mathcal{C}_{\tilde{\bm{u}}^{(n)}}$ encircle\fn{As briefly discussed in \cite{Kostov2}, the contours are in general complicated and the case-by-case analysis is necessary.} the Bethe roots $\bm{u}^{(n)}$ and $\tilde{\bm{u}}^{(n)}$ respectively.

So far, we have seen that the mixed correlators have simple expressions, which allow us to study the semi-classical limit with ease. The remaining class of three-point functions are the ones for which all the three operators are of the same type. We call such three-point functions ``unmixed''. It turns out that, in the case of the unmixed correlators, several different terms in the sum in \eqref{generalformula} contribute to the the highest power of $z_i$'s, and therefore the result cannot be simplified by the straightforward application of the aforementioned logic. In addition, the prediction from the semi-classical computation based on the coherent states (to be reported in \cite{KKN3}) does not take a form which can be readily obtained from the pDWPF. These two observations indicate  that the unmixed correlators are much more complicated objects. Nevertheless, studying such three-point functions is important for the following reason: The pDWPF is the quantity which describes the skew-symmetric product of two spin-chain state. Therefore, the fact that the mixed correlators can be reduced to the pDWPF suggests that such three-point functions are characterized essentially by the integrability governing the two-point function, which is already fairly well-understood. This in turn means that, in order to reveal the  genuine  ``integrability for the three-point functions'',  we do need to study the unmixed correlators, which cannot be simplified into the pDWPF.
\section{Monodromy relation\label{sec:mono}}
Based on the framework developed so far, we now derive  the 
second main result of this paper,  namely the nontrivial identities, to be called the monodromy relations,  satisfied by the two-point and the three-point functions with the monodromy operators inserted.  This  identity is a direct consequence of the two fundamental properties of the Lax operator, \ie  the ``unitarity'' and the ``crossing'',  and might provide a hint for the essence of the integrability of the correlation functions that we are eager to capture. 
\subsection{Monodromy relation for two-point functions\label{subsec:mono_2pt}}
First,  let us derive the  monodromy relation for the two-point functions using
the aforementioned   two basic properties of the Lax operator. 

\begin{figure}[t]
  \begin{center}
   \picture{clip,height=4cm}{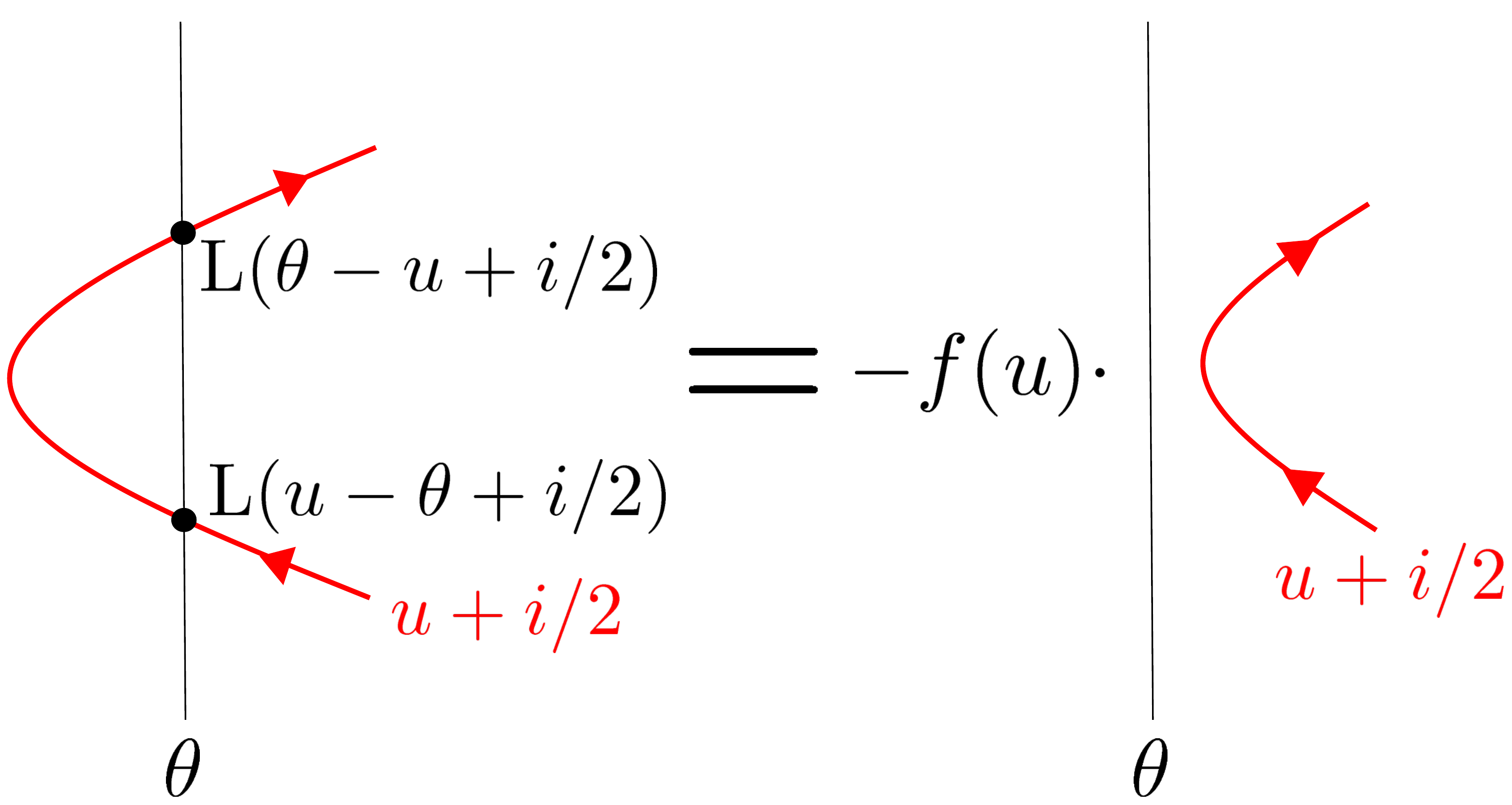}\\
   ``Unitarity'' relation.\\\vspace{5pt}
   \picture{clip,height=2.5cm}{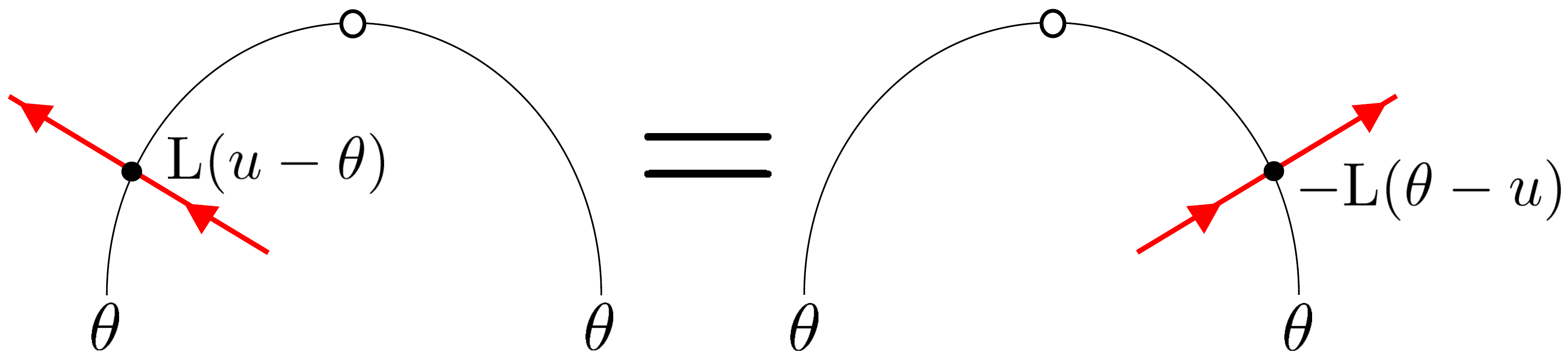}\\
   ``Crossing'' relation.
  \end{center}
\caption{The ``unitarity''  and the ``crossing'' relation of the Lax operator for the XXX spin chain. In both figures, the black line refers to  the spin space and the red line to the auxiliary space. Upper figure: A product of two Lax operators acting on the same spin space equals to the identity as shown in \eqref{unitarity}. Lower figure: A skew-symmetric product with a Lax operator insertion on one side is equivalent to the skew-symmetric product with a crossed Lax operator insertion on the other side as shown in \eqref{crossing2}.} 
\label{unitarity_crossing}
\end{figure}

The first  is the ``unitarity'' relation\fn{It  is also called the inversion identity.}. From the definition of the Lax operator \eqref{Laxdef}, one can straightforwardly check  the following identity:
\beq{
{\rm L}(\theta - u +i/2){\rm L}(u-\theta+i/2)=-f(u)\cdot \bm{1}\period \label{unitarity}
}
Here  the symbol ${\bm 1}$ denotes  the identity operator both for the spin and the auxiliary spaces and $f(u)$ is given by
\beq{
f(u)\equiv (u-\theta)^2 +1 \period
}
The relation \eqref{unitarity} is an analogue of the unitarity condition for the factorized S-matrices and can be understood pictorially as shown in the upper figure of \figref{unitarity_crossing}.

The second property is the ``crossing'' relation \eqref{crossing1}. What is important for the following discussions is that the crossed Lax operator $\mathcal{C}\circ {\rm L}(u)$ can be written alternatively as\fn{\label{fn:crossing}Written in terms of the R-matrix
 \beq{
 \left( R(u)\right)_{i_1 j_1}^{i_2 j_2}\equiv u\delta_{i_1}^{i_2}\delta_{j_1}^{j_2}+i\delta_{i_1}^{j_2}\delta_{i_2}^{j_1}\qquad (i_1,i_2,j_1,j_2=1,2)\comma
 }
which is related to the Lax operator by $R(u)={\rm L}(u+i/2)$, the equation \eqref{crossed2} takes the form of the crossing relation for the factorizable S-matrices,
 \beq{
 \sum_{i_1^{\prime}, i^{\prime}_2}-\left( \sigma_2\right)_{i_1 i^{\prime}_2} \left( R(u)\right)_{i^{\prime}_1 j_1}^{i^{\prime}_2 j_2} \left( \sigma_2\right)^{i^{\prime}_1 i_2} = \left(R(i-u)\right)_{i_1 j_1}^{i_2 j_2}\period
 }
This is the reason why we call $\mathcal{C}\circ {\rm L}(u)$ the crossed Lax operator.
}
\beq{
 \mathcal{C}\circ {\rm L}(u)= -{\rm L}(-u)\period\label{crossed2}
 }
 With this relation, the crossing relation \eqref{crossing1} takes the following form:
 \beq{
 \bra{\s}\left( {\rm L}(u-\theta)\ket{s_1}\otimes \ket{s_2}\right)=-\bra{\s}\left( \ket{s_1}\otimes {\rm L}(\theta-u)\ket{s_2}\right)\period\label{crossing2}
 }
 A pictorial representation of this relation is given in the lower figure of \figref{unitarity_crossing}.

\begin{figure}[t]
  \begin{center}
   \picture{clip,height=3.5cm}{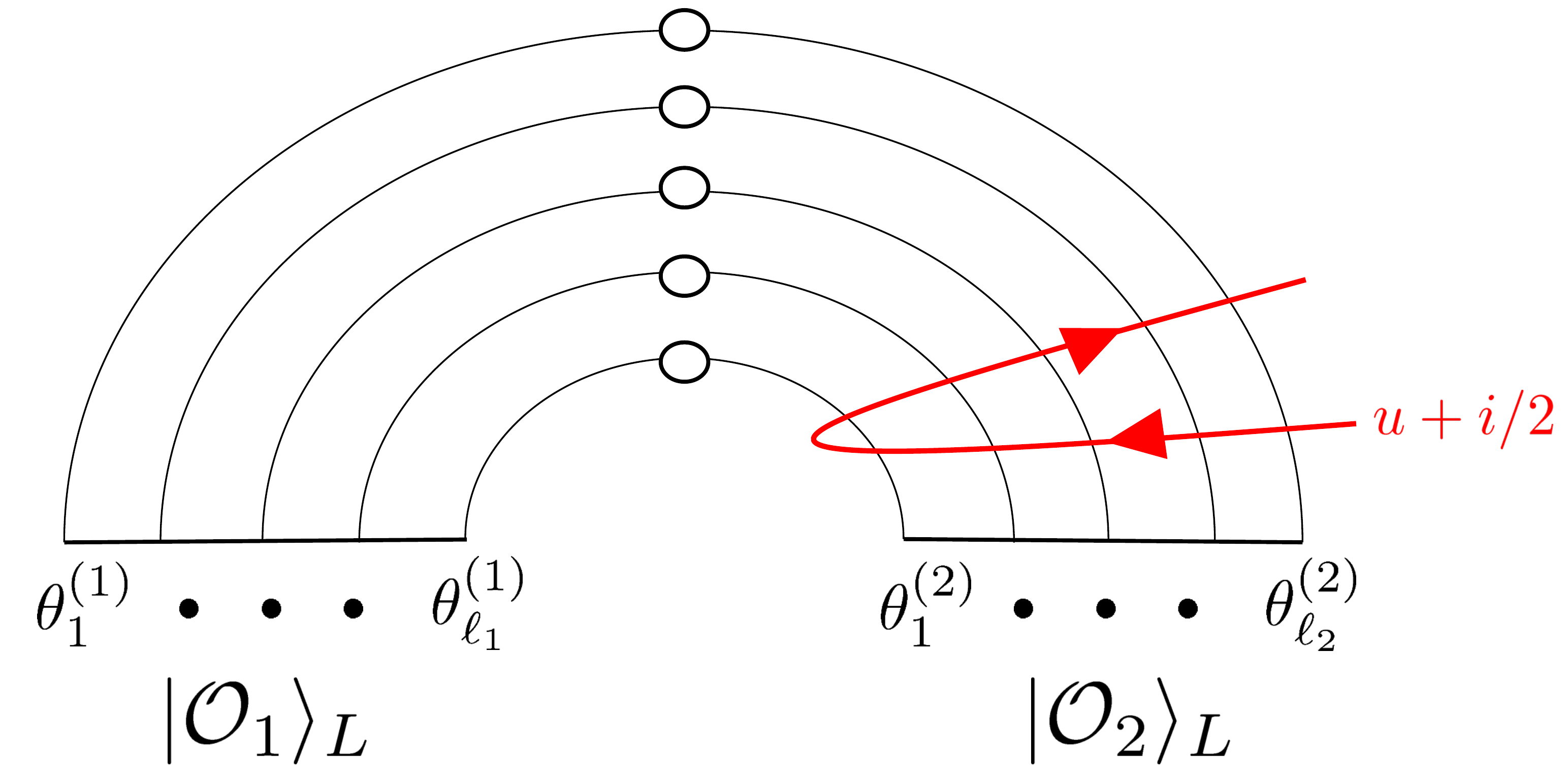}\\
    (a)
    \end{center}\vspace{5pt}
  \begin{minipage}{0.5\hsize}
  \begin{center}
   \picture{clip,height=3.5cm}{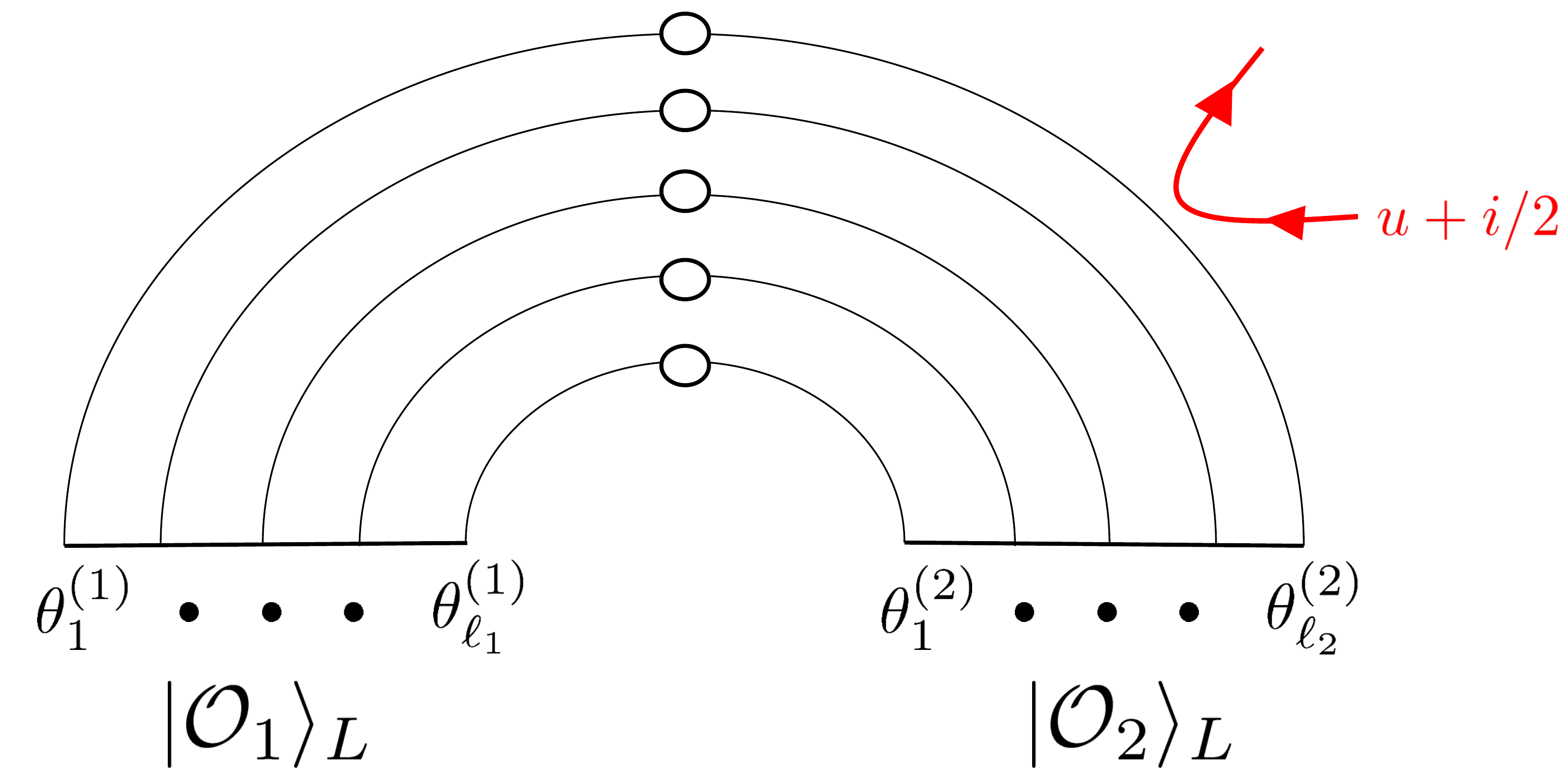}\\
  (b)
  \end{center}
 \end{minipage}
 \begin{minipage}{0.5\hsize}
  \begin{center}
   \picture{height=3.5cm,clip}{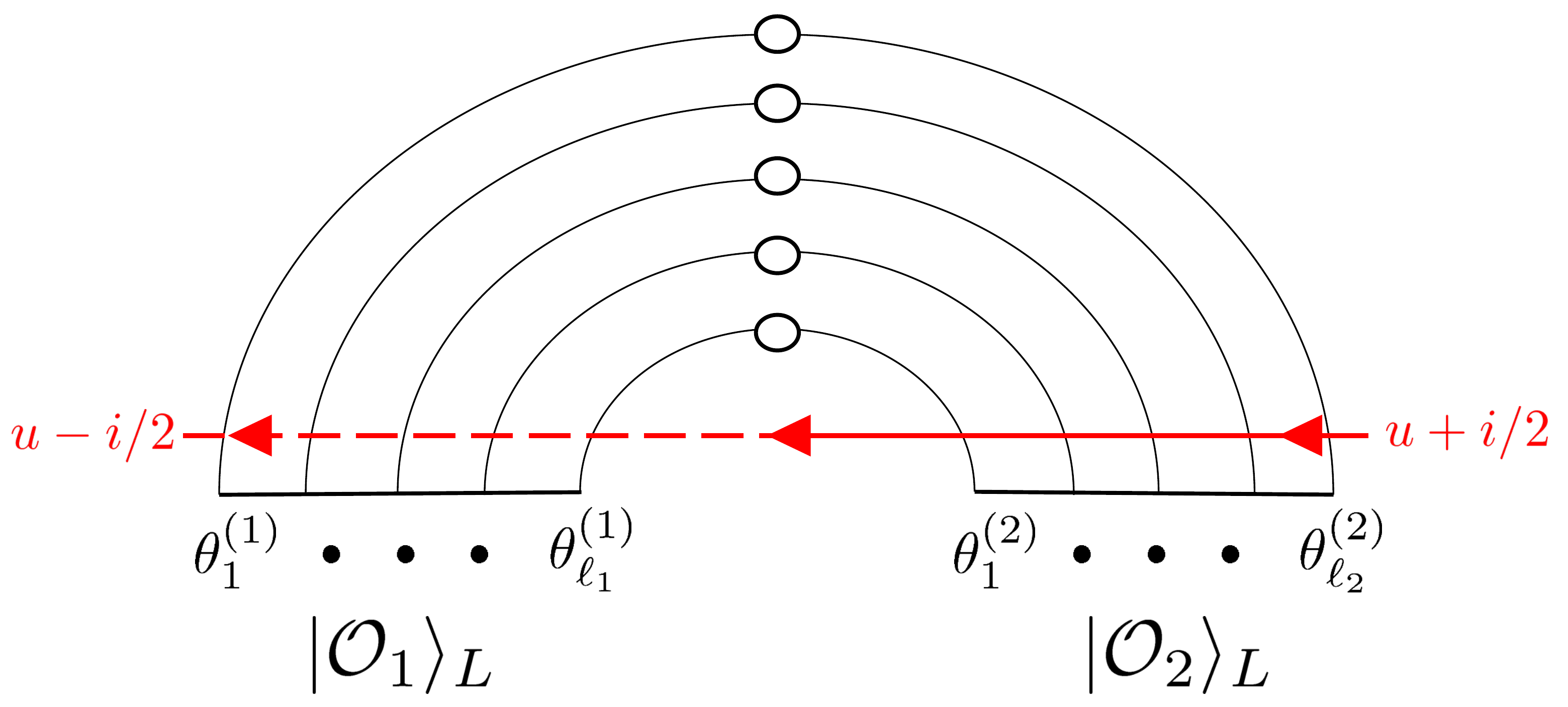}\\
  (c)
  \end{center}
 \end{minipage}
\caption{The derivation of the monodromy relation for the two-point function. The figure (a) describes the skew-symmetric product with the usual monodromy and the reverse-ordered monodromy, given in \eqref{omega2omega2}. By applying the unitarity relations, one can show that it is proportional to the skew-symmetric product without any monodromy insertions, which is given in \eqref{noomega} and depicted in the figure (b). On the other hand, if we apply the crossing relations repeatedly  to \eqref{omega2omega2}, we reach the right hand side of \eqref{omega+omega-}, which is shown in the figure (c). In the figure (c), the solid red line denotes the monodromy matrix whose argument is shifted by $+i/2$ whereas the dashed red line denotes the monodromy matrix whose argument is shifted by $-i/2$. The equivalence between the figures (b) and (c) is the monodromy relation for the two-point function given in \eqref{monod2_L}.} 
\label{2pt_monod}
\end{figure}
Making use of  these two properties\fn{For the moment, we only consider the SU(2)$_L$ sector since the generalization to the SU(2)$_R$ sector is straightforward.}, let us now derive the monodromy relation for the two-point functions . First, consider the following quantity, which is depicted in the figure (a) of \figref{2pt_monod}:
\beq{
\big<  \ket{\mathcal{O}_1}_L\comma \left( \overleftarrow{\Omega}_2 (-u+i/2)\right)_{ij}\Big( \Omega_2(u+i/2)\Big)_{jk} \ket{\mathcal{O}_2}_L  \big>\comma\label{omega2omega2}
}
where $i$, $j$ and $k$ are the indices for the auxiliary space, and $\Omega_n $ and  $\overleftarrow{\Omega}_n $ are the monodromy and the ``reverse-ordered'' monodromy\fn{Note that, owing to the relation \eqref{crossed2}, the reverse-ordered monodromy is equivalent to the monodromy which appeared in \eqref{eq-37}: $\overleftarrow{\Omega}_n(-u)=(-1)^{\ell_n}\mathcal{C}\circ \Omega_n(u)$.} for the operator $\mathcal{O}_n$, defined by
\beq{
\Omega_n(u+i/2)&\equiv {\rm L}^{(n)}_{1}(u-\theta^{(n)}_{1}+i/2)\cdots{\rm L}^{(n)}_{\ell_n}(u-\theta^{(n)}_{\ell_n}+i/2 )\comma\label{omegan}\\
\overleftarrow{\Omega}_n(-u+i/2)&\equiv {\rm L}^{(n)}_{\ell_n}(\theta^{(n)}_{\ell_n}-u+i/2 )\cdots{\rm L}^{(n)}_{1}(\theta^{(n)}_{1}-u+i/2)\period\label{revomegan}
}
Here ${\rm L}^{(n)}_k$ and $\theta^{(n)}_k$ respectively denote the Lax operator and the inhomogeneity parameter for the $k$-th site of the spin-chain state $\ket{\mathcal{O}_n}_{L}$, and $\ell_n$ is the length of the operator $\mathcal{O}_n$. Here again  the inhomogeneities are identified as
$\theta^{(1)}_k=\theta^{(2)}_{\ell-k+1}$,  as discussed already in section \ref{subsec:cutandsew}.
Using the unitarity relation \eqref{unitarity} repeatedly, we can show that \eqref{omega2omega2} is proportional to the skew-symmetric product without monodromy insertions, which is depicted in the figure (b) of \figref{2pt_monod}:
\beq{
\eqref{omega2omega2}=\delta_{ik}(-1)^{\ell} f_{12}(u)\big<  \ket{\mathcal{O}_1}_L\comma  \ket{\mathcal{O}_2}_L  \big>\comma\label{noomega}
}
where the prefactor $f_{12}(u)$ is given by
\beq{
&f_{12}(u)\equiv\prod_{k=1}^{\ell}\left((u-\theta^{(1)}_k)^2+1\right)
=\prod_{k=1}^{\ell}\left((u-\theta^{(2)}_k)^2+1\right)\period
}
Let us next apply the crossing relation to each Lax operator constituting $\overleftarrow{\Omega}_2$ in \eqref{omega2omega2}. Since the $k$-th site of the operator $\mathcal{O}_2$ is contracted with the $(\ell-k+1)$-th site of the operator $\mathcal{O}_1$, the Lax operator transforms under the application of the crossing relation as 
\beq{
{\rm L}^{(2)}_{k}(-u+\theta_k^{(2)})\to -{\rm L}^{(1)}_{\ell-k+1}(u-\theta_{\ell-k+1}^{(1)})\comma
}
where we used the identifications of  the inhomogeneity parameters \eqref{inhomoeq}. Thus, after the successive application of the crossing relation, we arrive at the following expression, which is depicted in the figure of \figref{2pt_monod}:
\beq{
\eqref{omega2omega2}=(-1)^{\ell}\big<\Big( \Omega_1^{-}(u)\Big)_{ij} \ket{\mathcal{O}_1}_L\comma  \Big( \Omega_2^{+}(u)\Big)_{jk} \ket{\mathcal{O}_2}_L  \big>\period\label{omega+omega-}
}
The superscripts $\pm$ on the monodromy operator denotes  the shift of the argument  $ \Omega^{\pm}(u)\equiv \Omega (u\pm i/2)$.

Then, by equating the right hand sides of \eqref{noomega} and \eqref{omega+omega-}, we obtain the monodromy relation for the two-point function:
\beq{
 \big<\Big( \Omega_1^{-}(u)\Big)_{ij} \ket{\mathcal{O}_1}_L\comma  \Big( \Omega_2^{+}(u)\Big)_{jk} \ket{\mathcal{O}_2}_L  \big>=\delta_{ik}\,f_{12}(u)\big<\ket{\mathcal{O}_1}_L\comma\ket{\mathcal{O}_2}_L \big> \period\label{monod2_L}
 }
One can write down a similar relation also for the SU(2)$_R$ chain as 
\beq{
 \big<\Big( \tilde{\Omega}_1^{-}(u)\Big)_{ij} \ket{\tilde{\mathcal{O}}_1}_R\comma  \Big( \tilde{\Omega}_2^{+}(u)\Big)_{jk} \ket{\tilde{\mathcal{O}}_2}_R  \big>=\delta_{ik}\,\tilde{f}_{12}(u)\big<\ket{\tilde{\mathcal{O}}_1}_R\comma\ket{\tilde{\mathcal{O}}_2}_R  \big> \comma\label{monod2_R}
 }
 where $\tilde{\Omega}_n(u)$ are the monodromy matrices for the SU(2)$_R$ chain and $\tilde{f}_{12}(u)$ is given in terms of the inhomogeneity for the SU(2)$_R$ chain $\tilde{\theta}^{(n)}_k$ by 
 \beq{
 \tilde{f}_{12}(u)\equiv \prod_{k=1}^{\ell}\left((u-\tilde{\theta}_k^{(1)})^2+1 \right)=\prod_{k=1}^{\ell}\left((u-\tilde{\theta}_k^{(2)})^2+1 \right)\period
 }
 
The monodromy relations \eqref{monod2_L} and \eqref{monod2_R} are the embodiment of the integrability for the two-point function. As the two-point function is determined by the spectrum of the operators, they should be essentially equivalent to the integrable structures already known in the spectral problem. However, it might be interesting to clarify the relation with the conventional formalism and ask if these new formalism helps  to deepen the  understanding of  the spectral problem. 

\subsection{Monodromy relation for three-point functions\label{subsec:mono_3pt}}
\begin{figure}[t]
  \begin{center}
   \picture{clip,height=3.5cm}{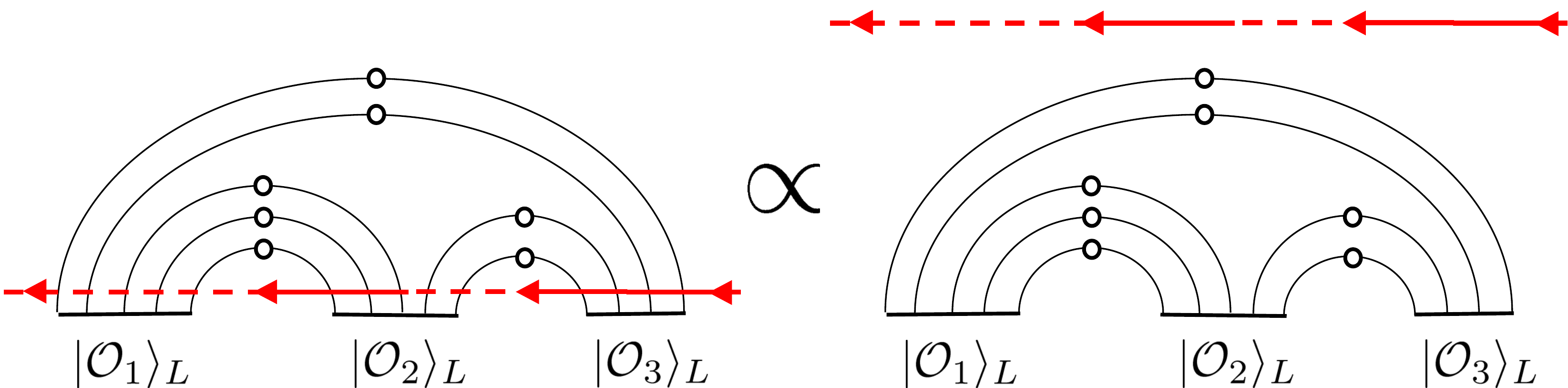}
  \end{center}
\caption{The monodromy relation for the three-point function \eqref{monod3_L}. The thick red line denotes a part of the monodromy matrix with a $+i/2$ shift of the spectral parameter and the dashed red line denotes a part of the monodromy matrix with a $-i/2$ shift of the spectral parameter.} 
\label{3pt_monod}
\end{figure}

Let us now turn to the three-point functions. As explained in section \ref{subsec:product}, the three-point functions are given by a product of two factors coming from the SU(2)$_L$ and the SU(2)$_R$ respectively and each factor is expressed in terms of the skew-symmetric products between sub-chains. Therefore, one can apply the unitarity \eqref{unitarity} and the crossing relation \eqref{crossing2} to each sub-chain and derive a nontrivial monodromy relation for the three-point functions. Although the essence of the derivation is entirely  similar to the one for the two-point function, for the three-point function there is a certain freedom in the form of the  monodromy relation which comes  from the choice of the shifts of the spectral parameter for the three monodromy matrices.  To give an intuitive picture of the monodromy relation, however, below we shall  exhibit a specific example which can be easily understood from  a figure \figref{3pt_monod} and  relegate the discussion of how the more general forms of the relation arise to  Appendix \ref{ap2}. 

Now for the SU(2)$_L$ sector a simple monodromy relation can be given in the form 
\beq{
\begin{aligned}
&\big<\Big(\Omega^{-}_1(u)\Big)_{ij}\ket{\mathcal{O}_1}_L\comma\Big(\Omega_2^{+|-}(u)\Big)_{jk}\ket{\mathcal{O}_2}_L \comma \Big(\Omega_3^{+}(u)\Big)_{kl}\ket{\mathcal{O}_3}_L\big>\\
&=\delta_{il}f_{123}(u)\big<\ket{\mathcal{O}_1}_L\comma\ket{\mathcal{O}_2}_L\comma \ket{\mathcal{O}_3}_L\big>\comma
\end{aligned}\label{monod3_L}
}
where $f_{123}(u)$ is defined by\fn{For a definition of $\ell_{ij}$, see \eqref{eq-20}.}
\beq{
f_{123}(u)\equiv \prod_{i=1}^{\ell_{31}}\left((u-\theta_i^{(1)})^2+1\right)
\prod_{j=1}^{\ell_{12}}\left((u-\theta_j^{(2)})^2+1\right)
\prod_{k=1}^{\ell_{23}}\left((u-\theta_k^{(3)})^2+1\right)\comma
 }
and $\Omega^{+|-}_2(u)$ denotes a product of the monodromy matrices on the left and the right sub-chains of $\mathcal{O}_2$ whose arguments are shifted by $+i/2$ and $-i/2$ respectively.
More specifically, the relevant monodromy matrices 
are given by 
\beq{
\begin{aligned}
&\Omega^{-}_1(u)={\rm L}^{-}_{1}(u-\theta^{(1)}_{1})\cdots {\rm L}^{-}_{\ell_{31}}(u-\theta^{(1)}_{\ell_{31}}){\rm L}^{-}_{\ell_{31}+1}(u-\theta^{(1)}_{\ell_{31}+1})\cdots {\rm L}^{-}_{\ell_1}(u-\theta^{(1)}_{\ell_1})\comma\\
&\Omega^{+|-}_2(u)={\rm L}^{+}_{1}(u-\theta^{(2)}_{1})\cdots {\rm L}^{+}_{\ell_{12}}(u-\theta^{(2)}_{\ell_{12}}){\rm L}^{-}_{\ell_{12}+1}(u-\theta^{(2)}_{\ell_{12}+1})\cdots {\rm L}^{-}_{\ell_2}(u-\theta^{(2)}_{\ell_2})\comma\\
&\Omega^{+}_3(u)={\rm L}^{+}_{1}(u-\theta^{(3)}_{1})\cdots {\rm L}^{+}_{\ell_{23}}(u-\theta^{(3)}_{\ell_{23}}){\rm L}^{+}_{\ell_{23}+1}(u-\theta^{(3)}_{\ell_{23}+1})\cdots {\rm L}^{+}_{\ell_3}(u-\theta^{(3)}_{\ell_3})\period
\end{aligned}\label{defshift}
}
For  the SU(2)$_R$ chain, the corresponding form of the monodromy relation can be written as 
\beq{
\begin{aligned}
&\big<\Big(\tilde{\Omega}^{-}_1(u)\Big)_{ij}\ket{\tilde{\mathcal{O}}_1}_R\comma\Big(\tilde{\Omega}_2^{+|-}(u)\Big)_{jk}\ket{\tilde{\mathcal{O}}_2}_R \comma \Big(\tilde{\Omega}_3^{+}(u)\Big)_{kl}\ket{\tilde{\mathcal{O}}_3}_R\big>\\
&=\delta_{il}\tilde{f}_{123}(u)\big<\ket{\tilde{\mathcal{O}}_1}_R\comma\ket{\tilde{\mathcal{O}}_2}_R\comma \ket{\tilde{\mathcal{O}}_3}_R\big>\comma
\end{aligned}\label{monod3_R}
}
 where $\tilde{f}_{123}(u)$ is defined by
\beq{
&\tilde{f}_{123}(u)\equiv \prod_{i=1}^{\ell_{31}}\left((u-\tilde{\theta}_i^{(1)})^2+1\right)
\prod_{j=1}^{\ell_{12}}\left((u-\tilde{\theta}_j^{(2)})^2+1\right)
\prod_{k=1}^{\ell_{23}}\left((u-\tilde{\theta}_k^{(3)})^2+1\right)\period
}
As in the SU(2)$_L$ sector, $\tilde{\Omega}^{+|-}_2(u)$ denotes a product of the monodromy matrices on the left and the right sub-chains whose arguments are shifted by $+i/2$ and $-i/2$ respectively.

Let us now discuss the implications of the typical monodromy relations of the form \eqref{monod3_L} and \eqref{monod3_R}. Firstly, the monodromy relations in general relate three-point functions of different spin-chain states and therefore can be regarded as a kind of Schwinger-Dyson equation. It would be extremely interesting if we could compute the three-point functions by directly solving these relations. Secondly, the relations imply the existence of infinite number of conserved charges together with  the existence of  the associated Ward identities. For instance,  by expanding \eqref{monod3_L} around $u=\infty$, at the leading order we obtain the usual Ward identities for the global SU(2)$_L$-symmetry of the form 
\beq{
&\big<S_{\ast}\ket{\mathcal{O}_1}_L\comma \ket{\mathcal{O}_2}_L\comma \ket{\mathcal{O}_3}_L\big>+\big<\ket{\mathcal{O}_1}_L\comma S_{\ast}\ket{\mathcal{O}_2}_L\comma \ket{\mathcal{O}_3}_L\big>+\big<\ket{\mathcal{O}_1}_L\comma \ket{\mathcal{O}_2}_L\comma S_{\ast}\ket{\mathcal{O}_3}_L\big>=0\comma \label{ward}
}
where $S_{\ast}$ are the global SU(2) generators and $\ast$ stands for $1$, $2$ or $3$. These global Ward identities are quite useful in fixing the kinematical dependence of the three-point functions, as  described in the Appendix \ref{ap}. Naturally it would be quite interesting and important to study  the non-trivial relations obtained at the sub-leading levels and see if we can exploit them to understand the structure of the three-point functions\fn{At the sub-leading order, \eqref{monod3_L} produces a set of non-trivial identities involving operators which act non-locally on the spin chains. These identities can be regarded as a sort of Yangian invariance for the three-point functions. Simlar relations are discussed in the context of the scattering amplitudes in \cite{spectralreg1,spectralreg2,spectralreg3,spectralreg4,Chicherin1,Chicherin2} and it would be interesting to clarify the connection.}. 

As for the importance of the monodromy relation, we already have a 
 supporting evidence from the strong coupling computation performed in \cite{KK3}.  In that analysis
  the three-point function in the SU(2) sector  was determined from the following relation for the monodromy matrices defined on the classical string world-sheet:
\beq{
\Omega_1(x)\Omega_2(x)\Omega_3(x)=\bm{1}\period\label{om1om2om3}
}
This relation, which is a direct consequence of the classical integrability of the string sigma model, is a clear manifestation of the integrability for the three-point function at strong coupling and was indeed an essential ingriedient in the  computation of  the three-point functions. The relations we derived here, \eqref{monod3_L} and \eqref{monod3_R}, can be regarded as the weak coupling counter-part of \eqref{om1om2om3} and its generalization.  The similarity becomes more apparent if we take the so-called semi-classical limit of the spin chain, in which the length of the chain and the number of the magnons are both large. To study the low energy excitation in this limit, we need to use the rescaled spectral parameter $u=\ell u^{\prime}$, and send $\ell$ to $\infty$ keeping $u^{\prime}$ finite. In terms of this rescaled parameter, the shifts of the spectral parameter in $\Omega_n^{\pm}$, $\Omega_2^{+|-}$ and so on become negligible. Furthermore, in this limit, the three-point function will be well-approximated by coherent states. Then the monodromy matrices, which are originally quantum operators acting on the spin chains, become classical. Therefore, in such a limit the relations \eqref{monod3_L} and \eqref{monod3_R} exactly take the same form as \eqref{om1om2om3}. As will be discussed in the forthcoming publication \cite{KKN3}, we can use \eqref{monod3_L} and \eqref{monod3_R} to directly study the semi-classical behavior of the three-point functions at weak coupling without relying on the explicit determinantal  expressions for the scalar products of the XXX spin chain.
\section{Discussions\label{sec:discuss}}
 In this paper, we proposed a novel way of understanding the tree-level three-point functions in the SU(2) sector. In the previous approaches, each operator was mapped to a single spin-chain state and the Wick contraction was interpreted as the scalar product of the spin chain. However, in order to study more general three-point functions, it is much more advantageous to associate a tensor product of two spin-chain states to each operator and express the Wick contraction as the overlap with the singlet state. Using this new formalism, we showed that a broader class of three-point functions, which we call mixed correlators, have simple determinant representation. Moreover, we derived nontrivial identities satisfied by the three-point functions with monodromy operators inserted. The identities can be regarded as the weak-coupling counterpart of the relation $\Omega_1 \Omega_2 \Omega_3 =1$, which played an important role in the computation at strong coupling. 

There are several future directions worth  exploring. One is to understand the loop corrections in this formalism. It was shown in \cite{Tailoring4,Fixing} that the loop corrections in the SU(2) sector can be neatly accounted for by the ingenious use of the inhomogeneities. It would be interesting if we can combine our formalism with the method in \cite{Tailoring4,Fixing}, and simplify and extend the computation at loop level. Another future direction is to study the ``unmixed'' correlators more in detail, for which we could not derive a simple expression. This is quite important since studying such correlators may help in revealing  the genuine characteristics of the three-point functions as discussed at the end of section \ref{subsec:det}.
It is also of importance to generalize this new formalism to other sectors, in particular to the non-compact sectors \cite{non-compact} and the sectors with fermions \cite{su(11)}. For the SL(2)-sector, one can indeed apply the idea developed in this paper and obtain useful results \cite{KKN4} for the three-point functions which are more general\fn{A different representation based on the separation of variables was obtained recently by \cite{Sobko}. In addition, the three-point functions for operators with spin were studied in detail in \cite{DO,Eden,AB} by using the operator product expansion of the four-point functions. It would be interesting to understand the relation with these works.} than the ones studied in \cite{non-compact}.

It would also be interesting if we can compute the three-point functions directly from the monodromy relations. Although it is currently not clear how to do it in the general setup, we can actually carry it out in the so-called semi-classical limit, in which the length of the operator and the number of the excitations are both large. As briefly mentioned at the end of section \ref{subsec:mono_3pt}, the monodromy relations in this limit take exactly the same form as the one at strong coupling $\Omega_1 \Omega_2 \Omega_3=1$. Then, one can study the semi-classical limit directly by generalizing the techniques developed in \cite{KK3} as will be discussed in the forthcoming publication \cite{KKN3}. Since such a method of computation does not rely on the explicit determinant expressions for the scalar products, it might be important for the higher-rank sectors, for which no useful determinant expressions are available except for special cases.

   Finally, it would be important to understand more conceptual aspects of our new formalism. As can be seen from \figref{2pt_comp} and \figref{3pt_comp}, the way we computed the three-point functions is analogous to the description of the interaction in the string field theory. If our formalism proves to be powerful also at the loop level, it may provide a useful framework to understand how the string field theory in the AdS background emerges from perturbative gauge theory\fn{Regarding this direction, there are several quite interesting works \cite{KR,stringbit}, which discuss the connection between the perturbative computation in the field theory and the string-field-theoretic formalism from a slightly different point of view.}.
\par\bigskip\noindent
{\large\bf Acknowledgment}\par\smallskip\noindent
S.K. would like to thank Y.~Jiang, I.~Kostov, D.~Serban and P.~Vieira for discussions.
The research of  Y.K. is supported in part by the 
 Grant-in-Aid for Scientific Research (B) 
No.~20340048, while that of T.N. is supported in part 
 by JSPS Research Fellowship for Young Scientists, from the Japan 
 Ministry of Education, Culture, Sports,  Science and Technology. The research of S.K. is supported by the Perimeter Institute for Theoretical Physics. Research
at Perimeter Institute is supported by the Government of Canada through Industry
Canada and by the Province of Ontario through the Ministry of Economic Development
and Innovation. S.K. acknowledges the kind hospitality of the Asian Pacific Center for Theoretical Physics and the Simons Center for Geometry and Physics during the completion of this work.
\appendix
\section{Kinematical dependence of the three-point function \label{ap}}
In this appendix,  we show that the kinematical dependence (\ie the dependence on the parameters $z_i$)  of the three-point functions can be completely determined by the invariance of the correlator under the symmetry group SO(4) $\cong$ SU(2)$_L\times$SU(2)$_R$  and the highest weight condition for the operators.  
This knowledge significantly simplifies  the calculation,  as elaborated  in subsection \ref{subsec:det}.

As usual, we concentrate on the \SUL sector. The key is the Ward identity  \eqref{ward} expressed using the the coherent state parametrization \eqref{3ptstates}. It is convenient to  remove the trivial overall factor from the states 
 in \eqref{eq-13} and consider
\begin{align}
|\hat{\mathcal{O}}_i\rangle_L := (1+|z_i|^2)^{L_i} |\mathcal{O}_i\rangle_L =e^{z_iS_-}|\boldsymbol{u}^{(i)}; \uparrow^{\ell_i} \rangle_L \comma \label{eq:normalize_z}
\end{align}
where $L_i=\ell_i/2-M_i$. Such a redefinition does not affect the Ward identity since $|\hat{\mathcal{O}}_i\rangle_L$ is related to $\ket{\mathcal{O}_i}_L$ simply by a multiplication of the scalar factor. It is important to note that the state $|\hat{\mathcal{O}}_i\rangle_L$  is independent of $\bar{z}_i$ as we have already implemented the highest weight condition. Hence, the remaining task is to determine the dependence on $z_i$.  

As is rather well-known, on such a coherent state representation, the 
\SUL generators have representations as differential operators. We can 
 easily show 
\begin{align}
&S_*|\hat{\mathcal{O}}_i \rangle_L= \rho_{z_i} (S_*) |\hat{\mathcal{O}}_i \rangle_L \comma \\
\rho_{z_i} (S_3)= L_i-z_i\frac{d}{dz_i}  \comma \ \ & \rho_{z_i}(S_+)= L_iz_i-\frac{z^2_i}{2}\frac{d}{dz_i}  \comma \ \ \rho_{z_i}(S_-)=\frac{d}{dz_i} \period \label{eq:diff_S}
\end{align}
For instance, the action on $S_3$ on the state $|\hat{\mathcal{O}}_i \rangle_L$ can be computed as 
\begin{align}
S_3 |\hat{\mathcal{O}}_i \rangle_L &= S_3 e^{z_i S_-} |\boldsymbol{u}^{(i)}; \uparrow^{\ell_i} \rangle_L  
= e^{z_i S_-} (e^{-z_iS_-} S_3 e^{z_i S_-}) |\boldsymbol{u}^{(i)}; \uparrow^{\ell_i} \rangle_L  \nn\\
&= e^{z_i S_-} (S_3 -z_iS_-) |\boldsymbol{u}^{(i)}; \uparrow^{\ell_i} \rangle_L  = \left(L_i -z_i {d \over dz_i} \right) |\hat{\mathcal{O}}_i \rangle_L 
\comma 
\end{align}
where $L_i$ is the eigenvalue of  $S_3$ on  $|\boldsymbol{u}^{(i)}; \uparrow^{\ell_i} \rangle_L $. 

Using this representation, Ward identity \eqref{ward} can be expressed as 
\begin{align}
\sum_{i=1}^3 \rho_{z_i}(S_*)  \big< |\hat{\mathcal{O}}_1 \rangle_L,|\hat{\mathcal{O}}_2 \rangle_L,|\hat{\mathcal{O}}_3 \rangle_L \big> =0 
\period
\end{align}
It is evident that this has  exactly the same form as the global conformal Ward identity for three-point functions in 2d CFT   if we identify $-L_i$ with the conformal dimensions. Thus, the $z_i$ dependence can be uniquely fixed \cite{BPZ} as  
\begin{align}
\big< |\hat{\mathcal{O}}_1 \rangle_L,|\hat{\mathcal{O}}_2 \rangle_L,|\hat{\mathcal{O}}_3 \rangle_L \big> \propto z_{21}^{L_{12}}z_{32}^{L_{23}}z_{13}^{L_{31}}\comma
\end{align} 
where $z_{ij}\equiv z_i-z_j$ and $L_{ij}\equiv L_i+L_j-L_k$. Therefore, the kinematical dependence of the three point function for the SU(2)$_L$ sector 
is given by the simple form 
\beq{
\begin{aligned}
\big< |\mathcal{O}_1 \rangle_L,|\mathcal{O}_2 \rangle_L,|\mathcal{O}_3 \rangle_L\big> \propto& \left( \frac{1}{1+|z_1|^2} \right)^{L_1} \left( \frac{1}{1+|z_2|^2} \right)^{L_2} \left( \frac{1}{1+|z_3|^2} \right)^{L_3}\\
&\times z_{21}^{L_{12}}z_{32}^{L_{23}}z_{13}^{L_{31}}\period
\end{aligned}\label{kineL}
}
Similarly, for the SU(2)$_R$ sector the result is 
\beq{
\begin{aligned}
\big< |\tilde{\mathcal{O}}_1 \rangle_R,|\tilde{\mathcal{O}}_2 \rangle_R,|\tilde{\mathcal{O}}_3 \rangle_R\big> \propto& \left( \frac{1}{1+|\tilde{z}_1|^2} \right)^{R_1} \left( \frac{1}{1+|\tilde{z}_2|^2} \right)^{R_2} \left( \frac{1}{1+|\tilde{z}_3|^2} \right)^{R_3}\\
&\times \tilde{z}_{21}^{R_{12}}\tilde{z}_{32}^{R_{23}}\tilde{z}_{13}^{R_{31}}\comma
\end{aligned}\label{kineR}
}
where $R_i$ is given by $\ell_i/2 -\tilde{M}_i$.
It is important to note  that the relations \eqref{kineL} and \eqref{kineR} take the following form in terms of the polarization spinors,
\beq{
\begin{aligned}
\big< |\mathcal{O}_1 \rangle_L,|\mathcal{O}_2 \rangle_L,|\mathcal{O}_3 \rangle_L\big> &\propto\langle\mathfrak{n}_1,\mathfrak{n}_2 \rangle^{L_{12}}\langle \mathfrak{n}_2,\mathfrak{n}_3 \rangle^{L_{23}}\langle \mathfrak{n}_3,\mathfrak{n}_1 \rangle^{L_{31}} \comma\\
\big< |\tilde{\mathcal{O}}_1 \rangle_R,|\tilde{\mathcal{O}}_2 \rangle_R,|\tilde{\mathcal{O}}_3 \rangle_R\big> &\propto\langle\tilde{\mathfrak{n}}_1,\tilde{\mathfrak{n}}_2 \rangle^{R_{12}}\langle \tilde{\mathfrak{n}}_2,\tilde{\mathfrak{n}}_3 \rangle^{R_{23}}\langle \tilde{\mathfrak{n}}_3,\tilde{\mathfrak{n}}_1 \rangle^{R_{31}}\comma
\end{aligned}
}
where $\langle\mathfrak{n},\mathfrak{m} \rangle\equiv \det \left(\mathfrak{n},\mathfrak{m} \right)$. This is precisely the structures observed in the computation at strong coupling \cite{KK3}.

It should be useful to make a small remark on the uniqueness of the kinematical dependence as  determined by the symmetry argument. Although the 
results \eqref{kineL} and \eqref{kineR} above for the ``SU(2) sector'' 
are unique, this is not true in the case of higher rank sectors. 
 For instance, in the SO(6) sector, the symmetry argument alone cannot fix the dependence completely  and there exist several possible R-symmetry tensorial structures. In such cases, the three-point function is given by a linear combination of such allowed  structures, whose coefficients depend on dynamics, 
for instance on 't Hooft coupling.  Indeed, for the SO(2,4) sector, the existence of a large number of  tensorial structures was found  in \cite{Spinning}.
\section{General form of the monodromy relation for three-point functions\label{ap2}}
In this appendix we briefly discuss  how more general forms  of the 
monodromy relations can be constructed. Below, for simplicity we shall suppress the inhomogeneity parameters  and consider the SU(2)$_L$ sector. 

The freedom in the form of the monodromy relation stems  from the simple 
 fact that  by making an arbitrary shift of $u$  the fundamental 
 unitarity relation \eqref{unitarity} can be rewritten as 
\begin{align}
 {\rm L}(-u+a) {\rm L}(u+b) &= -f(u+(b-a)/2)\cdot {\bf 1} \comma
\label{unitarity2} \\
a+b &= i \period
\end{align}
Now suppose we split each monodromy operator into left and the right parts, 
 similarly to  the case of $\Omega_2^{+|-}$ in \eqref{defshift}, in the form 
\begin{align}
\widehat{\Omega}_n(u) &\equiv \Omega_n^l(u+a_n^l) \Omega_n^r(u+a_n^r) \period
\end{align}
Then, by computing the three-point function 
$
\langle \left( \widehat{\Omega}_1(u) \right)_{ij}  \ket{\calO_1}_L, \left(
 \widehat{\Omega}_2(u) \right)_{jk}  \ket{\calO_2}_L, \left(\widehat{\Omega}_3(u) \right)_{kl}  \ket{\calO_3}_L \rangle  
$, using the crossing relations and Wick contractions, 
we easily find that the conditions for the coefficients $a_n^l$ and $a_n^r$ 
for which the unitarity relation \eqref{unitarity2} works
 to yield the result  proportional to  $\delta_{il}  \langle 
\ket{\calO_1}_L, \ket{\calO_1}_L, \ket{\calO_1}_L \rangle $  
are given by 
\begin{align}
a_2^l-a_1^r = a_3^l-a_2^r = a_3^r-a_1^l =i \period
\end{align}
For the simple example we discussed in section \ref{subsec:mono_3pt}, these relations are
satisfied with $a_1^l=a_1^r = -i/2, a_2^l =i/2, a_2^r=-i/2, 
a_3^l=a_3^r=i/2$.  In general, disregarding  a  common shift for  all the $a_n^{r,l}$,  there exist different monodromy relations which can be parametrized by two complex numbers.  At the moment, the meaning of this freedom is unclear to us. It might be a special feature of the tree-level relation. 
In any case, deeper understanding of the monodromy relation 
 is an important future problem.
\renewenvironment{thebibliography}{\pagebreak[3]\par\vspace{0.6em}
\begin{flushleft}{\large \bf References}\end{flushleft}
\vspace{-1.0em}
\renewcommand{\labelenumi}{[\arabic{enumi}]\ }
\begin{enumerate}\if@twocolumn\baselineskip=0.6em\itemsep -0.2em
\else\itemsep -0.2em\fi\labelsep 0.1em}{\end{enumerate} }

\end{document}